
\input psfig
\input mn.tex
\overfullrule=0pt




\def\kms{\;{\rm km\,s}^{-1}}

\def\frac#1#2{\textstyle {#1\over #2} \displaystyle}

\def\etal{{\rm et.al.}}

%
%
\def\spose#1{\hbox to 0pt{#1\hss}}
\def\lta{\mathrel{\spose{\lower 3pt\hbox{$\sim$}}
    \raise 2.0pt\hbox{$<$}}}
\def\gta{\mathrel{\spose{\lower 3pt\hbox{$\sim$}}
    \raise 2.0pt\hbox{$>$}}}
\def\today{\ifcase\month\or
 January\or February\or March\or April\or May\or June\or
 July\or August\or September\or October\or November\or December\fi
 \space\number\day, \number\year}
\newdimen\hssize
\hssize=8.4truecm  
\newdimen\hdsize
\hdsize=17.7truecm    


\newcount\eqnumber
\eqnumber=1
\def\chaphead{}
 
\def\new{\hbox{(\rm\chaphead\the\eqnumber)}\global\advance\eqnumber by 1}
 
\def\first{\hbox{(\rm\chaphead\the\eqnumber a)}\global\advance\eqnumber by 1}
\def\last#1{\advance\eqnumber by -1 \hbox{(\rm\chaphead\the\eqnumber#1)}\advance
     \eqnumber by 1}
 
\def\ref#1{\advance\eqnumber by -#1 \chaphead\the\eqnumber
     \advance\eqnumber by #1}
 
\def\nref#1{\advance\eqnumber by -#1 \chaphead\the\eqnumber
     \advance\eqnumber by #1}

\def\eqnam#1{\xdef#1{\chaphead\the\eqnumber}}
 


\newcount\tabnumber 
\tabnumber=1
\def\tabnew{\global\advance\tabnumber by 1}
\def\tabnam#1{\xdef#1{\chaphead\the\tabnumber}}


\newcount\fignumber 
\fignumber=1
\def\fignew{\global\advance\fignumber by 1}
\def\fignam#1{\xdef#1{\chaphead\the\fignumber}}


\pageoffset{-0.85truecm}{-1.05truecm}



\pagerange{}
\pubyear{version: \today}
\volume{}


\begintopmatter

\title{Dark Matter in Dwarf Spheroidals II: Observations and
       Modelling of Draco}

\author{J.\ Kleyna$^{1}$, M.I.\ Wilkinson$^{1}$,  N.W.\ Evans$^{2}$, G.\ Gilmore$^1$}

\vskip0.15truecm

\affiliation{$^1$ Institute of Astronomy, Madingley Rd, Cambridge CB3 0HA}

\vskip0.15truecm
\affiliation{$^2$ Theoretical Physics, Department of Physics, 1 Keble Road,
                 Oxford, OX1 3NP}

\shortauthor{J. Kleyna et al.}

\shorttitle{Dwarf Spheroidals II: Draco}



\abstract{We present stellar radial velocity data for the Draco dwarf
spheroidal (dSph) galaxy obtained using the AF2/WYFFOS instrument
combination on the William Herschel Telescope.  Our dataset consists
of 186 member stars, 159 of which have good quality velocities,
extending to a magnitude $V\approx 19.5$ with a mean velocity
precision of $\approx 2\,\kms$.  As this survey is based on a
high-precision photometric target list, it contains many more Draco
members at large radii. For the first time, this allows a robust
determination of the radial behaviour of the velocity dispersion in a
dSph.

We find statistically strong evidence of a rising velocity dispersion
consistent with a dark matter halo with gently rising rotation curve.
There is a $<2\sigma$ signature of rotation about the long axis,
inconsistent with tidal disruption being the source of the rising
dispersion. By comparing our dataset with earlier velocities, we find
that Draco probably has a binary distribution and fraction comparable
to those in the solar neighbourhood.

We apply a novel maximum likelihood algorithm and fit the velocity
data to a two parameter spherical model with an adjustable dark matter
content and velocity anisotropy.  Draco is best fit by a weakly
tangentially anisotropic distribution of stellar orbits in a dark
matter halo with a very slowly rising rotation law ($v_{\rm circ}
\propto r^{0.15}$).  We are able to rule out both a mass-follows-light
distribution and an extended halo with a harmonic core at the 2.5 to 
3$\sigma$ significance level, depending on the details of our
assumptions about Draco's stellar binary population. Our modelling
lends support to the idea that the dark matter in dwarf spheroidals is
distributed in the form of massive, nearly isothermal haloes.}


\maketitle  

\section{Introduction}
This paper describes observations and dynamical modeling of the Draco
dwarf spheroidal (dSph) galaxy.  Draco was the first dSph with
evidence of a large dark matter content.  Aaronson (1983) measured the
velocities of four carbon stars in Draco, inferring a mass far larger
than that corresponding to Draco's visible matter.  Subsequently,
Aaronson \& Olszewski (1987, 1988) obtained the velocities of $\sim
15$ more stars and computed a velocity dispersion of $\sim 10$ $\kms$
and a mass-to-light ratio of $>50\,M_\odot/L_\odot$.  These results
were confirmed by a much larger sample of 91 Draco stars by
Armandroff, Olszewski, \& Pryor (1995, hereafter AOP), and 17 largely
overlapping stars observed by Hargreaves et al. (1996).  Lokas (2001)
found that the dispersion profile of the inner part of Draco is
consistent with standard CDM halo models, but core data alone cannot
rule out mass-follows-light.  These studies yielded velocity
dispersions of 8.5 and 10.5 $\kms$ respectively.  The corresponding
mass-to-light ratios, assuming the standard mass-follows-light King
model (e.g., Binney \& Tremaine, chap. 4), were $90\pm15$ and
$145^{+116}_{-71}$ respectively.

However, these calculations depend on the assumption of a
mass-follows-light King model.  All the evidence from galactic
dynamics is that dark matter does not trace the luminous mass but
exists in extended halos (see e.g., Binney \& Tremaine 1987,
chap. 10).  Also, the King model assumes isotropic orbits and this
assumption hides the well-known degeneracy between mass and anisotropy.  A
tangentially anisotropic system, with orbits that become more circular
with increasing radius, may have a velocity dispersion that rises with
radius, effectively mimicking a massive halo.

In Sections 2 and 3 of the paper, we present the first Draco velocity
data at projected radii significantly outside the King core radius.
As our initial target list is based on accurate CCD data, rather than
photographic plate photometry, we are able to achieve a clean
separation between Draco's giant branch population and the Galactic
foreground out to distances as large as Draco's nominal tidal radius.
Consequently, these data contain many more stars at large radii than
previous studies. At projected radii $R>10^\prime$ (approximately the
King core radius), our sample has 63 stars, whereas the dataset of AOP
contains 13.  At $R>20^\prime$, our dataset contains 13 stars,
compared to only one star in AOP.  Furthermore, advances in
instrumentation have made the velocities of our dataset more accurate
than those previously published: our mean velocity uncertainty is 1.9
$\kms$, whereas the mean uncertainties of AOP and Hargreaves et
al. (1996) were 3.1 and $\ga 2$ $\kms$ respectively. In Section
4, we compare the velocities of stars in our dataset with the previous
data set of AOP and use the disagreement between the two to constrain
Draco's binary fraction.

Section 5 presents the dynamical modelling of the data. As an advance
beyond the standard mass-follows-light King models, we use a set of
spherically symmetric dSph models described in our companion paper
(Wilkinson et al. 2001, hereafter Paper I). The models have two free
parameters, the first being the orbital anisotropy and the second
being the dark matter content.  Some care is needed in comparing the
models to the data, as we must incorporate the effects of binary stars
on the velocity dispersion, as well as uncertainties in the
measurements.  Our model fitting represents the first attempt to probe
the spatial nature of a dSph dark matter halo and to disentangle the
effects of halo mass from those of orbital anisotropy.

\beginfigure*{\fignumber}
\fignam{\figstarcount}
\centerline{\psfig{figure=xyplot.ps,width=\hssize,angle=270}}
\smallskip\noindent
\caption{{\bf Figure \figstarcount.} Locations of observed members with good
velocities (open circles), probable members with low-quality
velocities (plus symbols) and non-members (dots).  Inner and outer
ellipses are the King core and tidal radii, as deduced from a fit to
the stellar surface number density (Irwin \& Hatzidimitriou 1995).}
\fignew
\endfigure

\section{Data}

Data were obtained from June 23 to June 26, 2000, at the William
Herschel Telscope using the AF2/WYFFOS fibre--fed spectrograph.  Sky
conditions were excellent over all four nights and no time was lost to
weather or instrument difficulties.  The one degree field of view of
this instrument is well--matched to the King tidal radius of Draco,
permitting all fields to be positioned on the same point at the centre
of Draco.  Typically, approximately 60 to 70 of the 101 functioning
fibres could be placed on target stars. Most of the remaining fibres
were allocated to sampling the sky spectrum.  Two fibre configurations
were observed per night. Typically, half an hour was lost to the
intervening fibre reconfiguration.  Usually, approximately seven
fields were obtained for each arrangement of fibres.  Fibres were
re-allocated in a manner that attempted to achieve comparable
signal-to-noise ($S/N$) for each star regardless of magnitude.  Stars
that are near the centre of Draco, or that had been observed
previously by other groups, were given a decreased weight when
allocating fibres.  Nevertheless, 62 of our final member stars have
previous velocity measurements.

Stars were selected from a prior $V$ and $I$ band CCD survey of Draco,
obtained with the 1.2m telescope at the Fred Lawrence Whipple
Observatory.  The 406 targets were chosen by selecting stars inside a
polygon tightly drawn around the giant branch branch, spanning a
magnitude range of $V\approx 17$ to $V\approx 19.8$.  The relatively
good quality of the CCD photometry used to compile the target list,
compared to the plate based photometry used as a basis for previous
radial velocity observations, allowed a high yield of member stars out
to Draco's tidal radius.  A total of 284 stars were observed, 203 of
which had good--quality velocities and 186 of which proved to be
probable Draco members.  Figure \figstarcount\ shows the positions of
observed stars, superimposed on the core and tidal radii of Draco;
open circles indicate members, plus symbols indicate probable
members with low-quality spectra, and dots indicate non-members.

The spectral range observed covered 8200 to 8800 \AA, corresponding to
the calcium triplet absorption line.  Observations were performed
using the 'echelle' grating with a resolution of 0.6
\AA\ per pixel.   Arc exposures using the ArNe lamp were taken 
every hour, so that each exposure was adjacent to at least one arc.
Additionally, twilight skies were taken each evening and morning to
obtain a velocity zero point using the solar spectrum.


\section{Data reduction}

Initial data reduction was performed using the WYFFOS-specific {\sc
wyffred} package inside IRAF.  {\sc wyffred} first identifies the
traces of individual spectra on the chip using a bright reference
image.  Next, it fits scattered light from the image with a product of
11th and 3rd order Chebyshev polynomials, along and across the
dispersion direction respectively.  {\sc wyffred} then uses the {\sc
identify} task to dispersion--correct the spectra.  For our spectra, a
typical wavelength solution involved fitting 21 lines from the ArNe
arc spectra with a fourth degree Chebyshev polynomial. Typically, the
median scatter was 0.02\AA, corresponding to 0.7 $\kms$.  The
wavelength calibration was extremely stable through the night. Hence,
each object exposure was calibrated using only its nearest--neighbour
arc exposure rather than with a linear combination of the two adjacent
arc exposures.

\subsection{Sky Subtraction}

The region of sky corresponding to the calcium triplet is very heavily
contaminated by sky emission lines.  The red--most line of the
triplet, at 8662 \AA, is masked by a complex of sky lines and was
removed from the analysis altogether.  The other two lines (at 8498
and 8542 \AA) are immediately adjacent to strong sky lines.  Thus,
accurate sky subtraction is extremely important to obtain a symmetric
peak in the cross-correlation of the object spectra with the template
spectrum.

{\sc wyffos} generates a composite sky spectrum for each exposure by
computing the mean of the spectra of the sky fibres, rejecting those
spectra that produce an excessive $\chi^2$.  Next, it uses relative
fibre throughputs obtained from a specially designated exposure to
subtract the appropriate sky component from each individual object
spectrum.  For this work, the standard procedure was modified slightly
to use the median of five throughput images rather than a single
image, in case a star should lie on a fibre in one of the throughput
images.

In theory (see e.g., Wyse \& Gilmore 1992), this procedure should
result in sky subtraction accurate to within one per cent.  In the
majority of cases, a good result was obtained. The residual of the sky
lines was approximately symmetrical about a mean of zero, indicating
that only the random Poisson noise element of the sky line remained
and that the mean component had been correctly subtracted.
Nevertheless, probably for reasons of non-Gaussian error
contributions, the relative fibre throughput was sometimes
miscomputed, resulting in incorrect sky subtraction and residual
systematic sky line contamination.  Because the sky lines contained of
the order of $10^3$ to $10^4$ counts per pixel, even a very small
systematic miscalculation of the sky could result in residual sky
lines large relative to the Ca absorption lines.  This residual
contamination was removed using a bootstrap procedure that is
described in the next subsection.

\subsection{Combining Spectra}

WYFFOS creates a set of multifibre spectra, each corresponding to a
single exposure.  Before cross-correlating the spectra with the
template, it is thus necessary to split each multifibre spectrum into
its constituent one--dimensional spectra, and then to combine all of
these one--dimensional spectra into a single object spectrum.  There
are a number of issues to consider when combining spectra. First, the
spectra may have come from different fibres, with differing
throughputs, so that spectra need to be weighted properly before
combination.  Next, the data have slightly different velocity
zero-points, necessitating conversion into the heliocentric frame
before combination.  Finally, the proper combination procedure must be
used to suppress cosmic rays and other noise, while discarding as
little data as possible.

The detailed procedure for combining spectra within IRAF is as
follows. The full multifibre spectra were placed into the heliocentric
frame using {\sc dopcor} and then were split into one--dimensional
spectra extending from 8470 to 8570 \AA\ using the task {\sc scopy},
thereby excising the contaminated red--most line of the Ca triplet.
These one--dimensional spectra were rescaled by the corresponding
fibre throughput, to make their flux values reflect the measured flux,
rather than the throughput--corrected flux that is the default for
{\sc wyffred}.

Next, the spectra were combined using the {\sc scombine} task.  For
combining, each spectrum was rescaled by its median flux, and then
these rescaled spectra were combined using the median. Median
combining was used only after considerable experimentation. For a
Gaussian data set, the median provides a $\sim 20$ per cent noisier
estimate of the true mean than the empirical mean of the data set.
Thus, using the median rather the the mean effectively discards one
third of the exposure time.  However, this defect is outweighed by the
superior non-Gaussian noise suppression of the median.  In particular,
the median strongly suppresses the residual cosmic rays left over from
the earlier data processing steps.  Once the data were combined, the
continuum spectrum was subtracted. As the wavelength range was only
100 \AA\ long by this point, the continuum was well represented by a
third degree polynomial.  Finally, the combined spectra were rebinned
to the same 0.01 \AA\ bin width.

As has been mentioned, the use of standard {\sc wyffred} background
subtraction did not suffice to remove the sky lines to within the
accuracy of Poisson statistics.  To circumvent this problem, we used
the sky spectrum itself as a sky line detection filter.  We convolved
a sky spectrum with each combined spectrum, computed the integrated
square of the convolution function, and then subtracted the amount of
sky that would minimise this integral.  In other words, residual sky
lines were removed by adding or subtracting sky from the combined
spectra until the convolution of the sky with the spectra was as small
as possible.  For most images, this made little difference, but for
some, it noticeably reduced residual sky lines.  The closest sky line
was $\sim 100$ $\kms$ removed from the mean velocity of Draco,
so this procedure did not introduce systematic changes in measured
stellar velocities. It did, however, reduce the incidence of 
double peaks and other artefacts in the final convolutions used to
obtain stellar velocities.

\beginfigure*{\fignumber}
\fignam{\figsamplespec}
\centerline{\psfig{figure=sample_spectra.ps,width=\hssize,angle=270}}
\smallskip\noindent
\caption{{\bf Figure \figsamplespec.} 
Top three panels: representative spectra of Draco member stars at
three increasingly faint magnitudes.  Fourth panel from top: the
synthetic template used for the cross correlation, containing the two
bluemost lines of the Ca triplet.  Bottom panel: typical sky
spectrum showing the strong sky lines.}
\fignew
\endfigure

\subsection{Cross Correlation}

To compute the velocity of stars from their spectra, we used the IRAF
{\sc fxcor} package to cross--correlate each stellar spectrum with a
synthetic template spectrum composed of the two blue--most lines of
the Ca triplet.  In an approximate match of the stellar Ca lines, the
template line at 8542 \AA\ had a 50 per cent larger amplitude than the
line at 8498 \AA, and both lines were Gaussian with a full width
half maximum of 2 \AA.

The cross correlation was performed over the entire 8470 - 8570 \AA\
range of the rebinned spectra.  The peak of the cross--correlation
function (CCF) was fit with a parabola spanning 20 rebinned pixels, or
2.0 \AA.  The details of the CCF fit had little influence on the final
velocity, but they did affect the Tonry \& Davis $R_{\rm TD}$ value
errors reported by {\sc fxcor}.  However, the velocity errors reported
by {\sc fxcor} were subsequently recalibrated by a multiplicative
constant so the latter effect was removed at the end.  The 159 stars
for which the CCF had a reasonably symmetrical unimodal peak were
accepted as valid members, with accurate velocities.  There were also
27 probable member stars with a poor quality CCF.  These stars were
excluded from our dynamics analyses because we did not judge their
velocities to be reliable without additional corroborating
observations.

Figure~\figsamplespec\ shows the spectral region of interest for three
representative spectra, after continuum subtraction.  Additionally,
the bottom panels of Figure~\figsamplespec\ show the template spectrum
and a typical sky spectrum.  The presence of the blue-shifted Ca lines
is clearly apparent in all of the spectra, including the faintest
spectrum at $V=19.29$.  This figure also shows the proximity of the
sky lines (bottom panel) to the shifted Ca lines, and the residuals of
the sky lines in the final spectra.

\beginfigure*{\fignumber}
\fignam{\figerrors}
\centerline{\psfig{figure=triple_error.ps,width=\hssize,angle=270}}
\smallskip\noindent
\caption{{\bf Figure \figerrors.} 
Top panel: empirical velocity errors as a function of $S/N$ relative
to the $S/N$ of a $V=17$ star observed for one hour.  These errors
were obtained by dividing the spectra of bright stars into two
batches, using these batches to obtain two velocities, and dividing
the difference between the two velocities by $\sqrt{2}$ to give the
velocity error at each point.  Centre panel: ratio of the above
empirical errors to the nominal IRAF velocity uncertainty.  The dotted
line at 0.35 is the number by which the IRAF errors must be rescaled
to give the expected $\chi^2$ per degree of freedom.  Bottom panel:
errors predicted by Monte Carlo simulations, and the rescaling (dotted
line) necessary to make these errors agree with the empirical errors
of the top panel.  As the Monte Carlo errors produce 4 to 5$\sigma$
outliers when used to predict the observed errors, so we elected to
use the well-behaved rescaled IRAF errors of the centre panel as the
basis of our error model.}
\fignew
\endfigure

\subsection{Error Analysis}

Accurate error estimation is perhaps the most difficult component of
the data analysis.  Unlike previous authors (Olszewski et al. 1996), we
have only one epoch of data, preventing us from directly measuring the
velocity uncertainties though $\chi^2$ discrepancies of measurements
taken at different times.

We employed two approaches to estimating the velocity errors.  First,
we recombined the brighter objects ($V<18.5$) into two final `split'
spectra, one composed of the odd numbered exposures and the other of
the even exposures.  We then cross-correlated these two spectra with
the template separately and used the discrepancy between the two
velocities obtained to estimate the true error $\sigma_{\rm obs}$ of
the measurement (Figure \figerrors, top panel).  This error in turn
was used to obtain a rescaling factor to apply to the Tonry \& Davis 
$R_{\rm TD}$ value errors reported by {\sc fxcor} (Figure \figerrors, centre
panel).  We found that we needed to rescale the {\sc fxcor} errors by
0.35 (Figure \figerrors, centre panel, dotted line) to produce the
desired $\chi^2$ per degree of freedom of 1.0.  From the centre panel
of Figure \figerrors, it is apparent that, after rescaling
$\sigma_{\rm obs}$ by 0.35, only one $3\sigma$ outlier remains,
consistent with a well-behaved Gaussian error distribution with no
extreme outliers.

Our second approach for estimating velocity errors was to build a
Monte Carlo library of simulated observations, parameterised by
stellar magnitude and effective (fibre throughput adjusted) exposure
time.  We then created an interpolation table of velocity error as a
function of exposure time and magnitude.  Next, we compared the
discrepancies observed between the split spectra with those predicted by
the interpolation table and found them to agree to within 10 per cent
on average.  However, the $\sigma_{\rm obs}$ of a small number of the
split spectra disagreed by 4 to 5$\sigma$ with the value predicted by
the table (Figure \figerrors, bottom panel, outliers).

On the basis of this error analysis, we elected to reject the Monte
Carlo approach, and model our errors by rescaling the {\sc fxcor}
errors by 0.35. This rescaling was used for the entire data set
including those stars too faint to be split into two usable spectra.
Of the 284 stars for which we obtained velocities, 203 had good
quality velocities as judged by a clear peak in the CCF.  Of these 203
stars, 159 were Draco members.  An additional 27 stars were probable
members, but the CCF was multi-modal or highly asymmetric.  Figure 4
shows a histogram of stellar velocities of the stars with valid
velocities, extending to the velocity of the first star judged to be a
non-member.  The separation between Draco members and Galactic
contaminant stars is very clear, and there is little likelihood of
non-Draco stars being included in the sample.

It is apparent that many of the stars have a low Tonry \& Davis
(1979) $R_{\rm TD}$ value; 10 stars of the 159 star sample have
$R_{\rm TD}<2$.  Tonry \& Davis (1979) point out that the velocity
error formula $\sigma_v\propto (1+R_{\rm TD})^{-1}$ begins to break
down for small values of $R_{\rm TD}$, and attribute this breakdown to
the selection of false CCF peaks.  This may be worrisome, because our
IRAF-based velocity errors depend on the validity of this error
scaling.  However, this problem is mitigated in several ways.  First,
we reject stars with an obviously multi-modal CCF.  Next, our
Monte-Carlo simulations indicate that reliable velocities are possible
even for the faint stars of our sample, demonstrating that wildly
spurious CCF peaks occur infrequently.  Finally, we note that the
effect of additional velocity errors is milder than the effect of
additional binaries, which we model extensively. For instance, in our
following dynamical analyses, we assume binary distributions that have
between 5 and 15 binaries moving faster than 10 $\kms$, and demonstrate
that these have little effect on our dynamical inferences.
Nevertheless, in the following sections, we repeat some of our
analyses using subsamples of stars with $R_{\rm TD}>5$ and
$R_{\rm TD}>7$ to demonstrate that our dynamics conclusions 
remain unchanged if we limit our analysis to high quality spectra.

\beginfigure*{\fignumber}
\fignam{\figvelhist}
\centerline{\psfig{figure=hist.ps,width=\hssize,angle=270}}
\smallskip\noindent
\caption{{\bf Figure \figvelhist.} 
Histogram of stellar velocities of observed stars, including the
closest member of Galactic contaminant population at $\sim 220$
$\kms$.}
\fignew
\endfigure

\subsection{Velocity Results}

The mean velocity of the 159 star sample, including the heliocentric
correction but before a zero-point velocity correction, is $-290.6\pm
0.8$ $\kms$.  This agrees with the the values $-293.3\pm 1.0$ $\kms$
measured by AOP, and $-293.8\pm2.7$ measured by Hargreaves
\etal\ (1996) at the $2.1\sigma$ level.  However, our zero point
correction obtained from the twilight sky frames produces a corrected
mean velocity equal to $-287.1\pm0.9$ $\kms$, which is not in agreement
with these previous data.  The source of this $4.6\sigma$ discrepancy
is unknown.

In lieu of using our questionable velocity zero point, we
instead elect to use the 61 stars common to our data 
set and that of AOP to calibrate the mean velocity our data.  
The mean (median) velocity difference between the two
data sets is -1.22 (1.36) $\kms$; accordingly, our
reported velocities are adjusted by a zero point
correction of +1.22 $\kms$.  This adjustment leaves
the mean velocity of our entire data set, including stars
not present in previous data, in $1.1\sigma$ agreement
with the entire AOP data set.

The twilight sky images, although apparently unsuitable for a precise
zero point calibration, did demonstrate that we maintained stable
relative velocities throughout the observing run.  The RMS scatter of
the nightly mean velocity of the solar Ca lines was $0.34$ $\kms$ across
the four observing nights.

Table 1 gives the 159 member stars stars with good velocities.  The
first column is our internal target identification number. For stars
with velocities also measured by AOP, the second column gives their
primary name for this star.  The other columns provice J2000 right
ascension and declination, velocity, velocity uncertainty, and Tonry
\& Davis (1979) $R_{\rm TD}$ value of the CCF.  The velocity has been
adjusted by $+1.22$ $\kms$ to match the mean velocith of AOP. Table 2
gives the remaining 27 probable member stars.  These stars have a
multi-modal or strongly asymmetric CCF, and were not included in our
subsequent dynamics analyses.  The one star ({\it ID}$=273$) of these
27 that also appears in the previously published data set of AOP is in
$2.9\sigma$ disagreement with this previous published velocity,
supporting our decision to reject it based on its CCF.

\begintable*{\tabnumber}
\tabnam{\stars}
\tabnam{\stars}
\caption{{\bf Table \stars} Observed member stars - {\it ID} is the 
identification number in our target catalogue; {\it ID}$_{\rm AOP}$ is
the designation of the object in the previous data set of Armandroff,
Olszewski, \& Pryor (1995); $V$ is the $V$-band magnitude; {\it RA}
and {\it Dec} are the J2000 coordinates; $v$ and $\sigma_v$ are
measured heliocentric velocity and its uncertainty; and $R_{\rm TD}$ is the
Tonry \& Davis (1979) $R$ value of the cross-correlation function.}
\halign{
#&\quad#&\enskip#&\quad#&\enskip#&\enskip#&\quad#&\enskip#&
\enskip#&\quad#&\quad#&\quad#
&\qquad#& 
#&\quad#&\enskip#&\quad#&\enskip#&\enskip#&\quad#&\enskip#&
\enskip#&\quad#&\quad#&\quad#
\cr
\noalign{\hrule}
\noalign{\vskip0.2truecm}
{\it ID}& {\it ID$_{\rm AOP}$} & \hfil{\it V}\hfil
& \multispan3{\it \hfill RA \hfill} 
& \multispan3{\it \hfill  Dec \hfill } & \hfil $v$ \hfil & 
\hfil $\sigma_v$  \hfil & \hfil $R_{\rm TD}$ \hfil
&\hfil \quad \hfil&
{\it ID}& {\it ID$_{\rm AOP}$} & \hfil{\it V}\hfil 
& \multispan3{\it \hfill RA \hfill} 
& \multispan3{\it \hfill  Dec \hfill } & \hfil $v$ \hfil & 
\hfil $\sigma_v$  \hfil & \hfil $R_{\rm TD}$  \hfil \cr
\noalign{\vskip0.2truecm}
14&&18.26&17&19&16.94&+58&15&50.8&-290.17&1.81&5.2&&154&G&17.69&17&20&27.42&+57&56&52.4&-291.41&0.80&13.0\cr
16&13784&17.22&17&21&02.26&+58&15&38.3&-277.35&0.89&12.2&&158&&19.20&17&20&11.09&+57&56&45.7&-278.82&2.27&3.7\cr
23&&18.43&17&20&14.55&+58&12&51.4&-301.23&2.11&4.2&&160&273&18.37&17&19&50.11&+57&56&40.7&-281.15&2.46&3.7\cr
29&&18.49&17&18&05.77&+58&11&22.0&-286.57&1.91&4.3&&163&&18.43&17&22&11.25&+57&56&33.2&-311.32&1.92&4.6\cr
39&&19.35&17&18&45.87&+58&07&32.7&-265.70&4.57&1.8&&166&473&17.61&17&19&36.06&+57&56&29.0&-288.50&0.68&16.9\cr
43&&18.41&17&19&53.89&+58&06&44.1&-277.69&1.47&7.0&&168&&19.01&17&20&14.84&+57&56&25.6&-286.26&3.36&2.3\cr
47&&18.24&17&17&58.39&+58&05&58.6&-297.89&3.37&2.4&&171&&18.61&17&21&17.93&+57&56&21.4&-291.42&1.76&5.4\cr
48&&19.49&17&20&32.16&+58&05&39.9&-280.82&3.57&1.9&&173&&18.66&17&20&40.35&+57&56&18.2&-292.09&1.46&6.9\cr
49&&18.87&17&20&23.76&+58&05&24.2&-309.12&2.50&3.4&&174&&18.40&17&19&53.52&+57&56&16.5&-295.86&1.38&7.5\cr
50&&19.12&17&19&37.13&+58&05&20.1&-274.43&2.59&3.7&&176&L&18.32&17&20&17.07&+57&56&12.3&-295.03&1.14&9.4\cr
52&&19.13&17&21&05.85&+58&04&55.0&-290.10&3.88&1.6&&177&&19.52&17&22&06.60&+57&56&10.4&-302.82&4.23&2.0\cr
55&&17.32&17&17&38.58&+58&04&38.4&-267.13&0.81&12.4&&184&3309&18.37&17&21&03.26&+57&55&59.4&-290.53&0.84&13.2\cr
57&XI-13&17.23&17&18&52.15&+58&04&13.0&-295.45&0.75&15.8&&185&&18.18&17&17&55.41&+57&55&59.0&-291.46&1.27&7.8\cr
62&XI-7&17.44&17&19&12.30&+58&02&57.8&-291.15&1.11&10.0&&186&&18.95&17&19&26.83&+57&55&51.8&-279.99&1.77&5.0\cr
63&&18.91&17&19&36.67&+58&02&50.4&-290.18&2.44&3.6&&187&&19.18&17&21&41.83&+57&55&50.9&-283.17&2.86&2.7\cr
66&&19.25&17&20&56.62&+58&02&35.2&-283.19&2.53&3.5&&189&20083&17.83&17&18&57.81&+57&55&46.3&-296.67&1.22&8.4\cr
68&&19.17&17&21&22.32&+58&02&15.7&-307.75&2.30&4.4&&192&3116&18.57&17&19&22.97&+57&55&30.8&-280.06&1.68&5.7\cr
69&17557&18.24&17&21&04.89&+58&02&01.2&-298.34&1.34&7.7&&197&&19.06&17&20&24.16&+57&55&15.6&-289.69&2.10&4.2\cr
70&&18.34&17&19&21.85&+58&01&52.3&-295.77&1.22&8.5&&198&286&17.81&17&19&45.19&+57&55&14.4&-297.05&0.89&12.8\cr
71&17871&17.26&17&18&08.10&+58&01&46.7&-310.44&0.50&21.8&&199&&19.84&17&18&24.05&+57&55&08.1&-303.03&5.46&1.4\cr
74&&19.02&17&20&01.59&+58&01&33.7&-295.27&1.21&8.4&&201&&18.36&17&19&57.37&+57&55&04.8&-306.26&1.88&5.5\cr
76&&19.77&17&22&02.34&+58&01&17.4&-296.01&5.35&1.4&&205&&19.10&17&20&44.74&+57&54&59.9&-293.71&3.94&2.2\cr
79&&18.78&17&20&18.03&+58&01&12.1&-287.73&2.68&3.5&&206&&19.25&17&21&35.42&+57&54&59.5&-296.36&2.90&2.4\cr
83&&19.08&17&21&37.81&+58&00&52.5&-282.70&3.37&2.5&&208&297&17.85&17&19&41.21&+57&54&56.8&-285.52&0.82&14.0\cr
84&&18.79&17&18&35.67&+58&00&50.1&-289.61&3.30&2.4&&212&&18.73&17&20&25.02&+57&54&50.8&-277.27&1.40&6.4\cr
85&&18.40&17&19&21.76&+58&00&40.6&-275.48&1.49&7.0&&216&3297&17.89&17&20&53.79&+57&54&40.8&-284.47&1.09&9.8\cr
86&&18.85&17&20&15.27&+58&00&35.1&-296.57&1.61&6.5&&219&&18.31&17&18&44.56&+57&54&38.3&-284.47&1.24&8.9\cr
92&&19.09&17&19&07.79&+58&00&19.8&-287.45&3.26&3.0&&220&&18.96&17&19&10.33&+57&54&37.7&-295.81&2.44&3.4\cr
93&&19.13&17&22&06.12&+58&00&12.0&-289.21&3.19&2.3&&222&&18.34&17&19&57.73&+57&54&35.3&-296.41&1.04&10.4\cr
96&581&17.73&17&20&34.81&+57&59&56.9&-297.70&0.70&17.0&&223&&18.83&17&20&07.52&+57&54&32.8&-299.08&1.89&4.8\cr
98&3363&17.39&17&20&47.82&+57&59&55.6&-294.30&0.91&12.7&&224&&18.09&17&19&17.65&+57&54&32.4&-289.52&1.36&7.7\cr
99&&18.37&17&18&41.42&+57&59&52.1&-262.99&1.85&5.7&&230&490&17.66&17&19&40.04&+57&54&24.9&-304.72&0.74&15.2\cr
100&3361&17.58&17&20&52.10&+57&59&47.8&-294.43&0.90&12.5&&233&IX-18&17.77&17&18&57.66&+57&54&14.3&-278.58&1.14&8.8\cr
101&&18.40&17&19&20.05&+57&59&43.1&-306.59&2.13&4.3&&240&IV-18&17.47&17&22&10.75&+57&53&57.4&-324.43&0.63&16.2\cr
104&18281&18.11&17&21&56.99&+57&59&33.9&-314.16&5.37&1.6&&245&K&18.23&17&19&55.86&+57&53&48.8&-293.61&1.43&7.3\cr
105&449&17.51&17&19&51.86&+57&59&17.9&-302.99&0.76&15.2&&248&H&18.28&17&20&15.77&+57&53&43.5&-298.30&1.30&8.2\cr
106&X-11&17.18&17&19&10.88&+57&59&17.2&-297.17&0.72&15.4&&250&&18.65&17&21&49.22&+57&53&37.2&-291.54&2.04&4.7\cr
107&&18.54&17&19&56.45&+57&59&16.1&-294.55&3.06&2.9&&253&361&17.56&17&20&34.22&+57&53&32.1&-286.16&0.94&12.5\cr
109&18723&17.20&17&18&42.30&+57&59&10.0&-290.49&0.77&14.2&&259&&18.00&17&20&17.01&+57&53&12.4&-290.06&1.14&10.2\cr
111&437&17.69&17&20&17.06&+57&59&02.1&-304.97&1.37&7.5&&262&IV-20&17.39&17&22&13.68&+57&53&06.7&-300.23&0.77&14.6\cr
112&&18.40&17&17&38.60&+57&58&54.1&-285.65&3.42&2.5&&265&&18.68&17&19&57.10&+57&52&58.1&-272.54&1.68&5.7\cr
115&3053&17.46&17&19&35.51&+57&58&46.6&-282.60&1.04&11.3&&270&&19.04&17&21&13.88&+57&52&46.0&-280.86&3.38&2.7\cr
116&461&17.26&17&19&42.44&+57&58&37.7&-293.36&3.07&2.4&&272&&18.48&17&19&56.66&+57&52&43.0&-295.53&1.38&8.3\cr
117&&19.07&17&21&36.85&+57&58&37.1&-303.84&2.97&2.3&&275&IX-5&17.53&17&19&13.51&+57&52&33.6&-286.74&0.99&11.4\cr
121&&18.78&17&20&41.10&+57&58&25.7&-295.70&2.20&3.8&&278&&18.54&17&19&56.98&+57&52&23.1&-272.94&1.93&5.1\cr
122&&18.48&17&20&34.12&+57&58&22.7&-296.04&1.81&5.4&&279&&18.29&17&21&14.27&+57&52&22.5&-290.85&1.34&7.9\cr
123&&18.46&17&20&10.54&+57&58&20.9&-272.80&1.25&8.1&&284&21456&17.15&17&18&10.43&+57&52&09.2&-306.42&0.55&19.3\cr
125&3339&18.11&17&20&53.77&+57&58&12.6&-288.64&1.36&8.2&&285&343&17.63&17&20&13.59&+57&51&59.4&-294.22&1.11&10.1\cr
129&&19.32&17&20&33.53&+57&58&06.2&-286.38&3.40&1.8&&286&350&18.35&17&20&20.44&+57&51&58.4&-292.34&1.10&9.8\cr
130&427&18.14&17&20&31.22&+57&58&05.5&-302.57&1.33&7.7&&287&&18.65&17&20&56.73&+57&51&57.5&-292.73&2.43&3.9\cr
134&11&17.64&17&20&05.74&+57&57&52.7&-284.28&0.82&13.9&&288&&19.13&17&19&13.94&+57&51&55.2&-298.03&2.31&3.4\cr
136&&19.29&17&18&25.19&+57&57&45.6&-302.00&2.66&2.7&&291&&18.50&17&18&30.61&+57&51&47.0&-283.07&2.46&3.9\cr
138&&18.90&17&22&43.71&+57&57&30.8&-279.77&1.76&5.8&&295&506&18.04&17&19&53.13&+57&51&38.0&-303.70&1.06&8.8\cr
141&&19.06&17&22&18.65&+57&57&19.4&-281.89&2.49&3.5&&297&&18.82&17&20&02.03&+57&51&30.5&-312.78&1.71&5.6\cr
145&3316&17.40&17&21&03.60&+57&57&08.3&-287.75&0.69&16.2&&299&3150&17.38&17&19&36.37&+57&51&25.1&-280.42&0.77&14.5\cr
146&&18.60&17&20&24.57&+57&57&08.2&-290.76&1.51&6.6&&301&3255&18.26&17&20&39.90&+57&51&15.4&-282.95&1.31&8.1\cr
149&22&18.22&17&20&01.66&+57&57&04.6&-296.58&1.18&8.9&&302&3267&17.92&17&20&51.00&+57&51&09.8&-282.19&1.02&8.9\cr
151&45&17.76&17&19&57.99&+57&56&58.4&-292.71&1.11&10.1&&306&&19.41&17&19&23.31&+57&51&01.4&-289.71&3.04&2.6\cr
\noalign{\vskip0.3truecm}
\noalign{\hrule}
}
\tabnew
\endtable

\begintable*{\tabnumber}
\tabnam{\stars}
\caption{{\bf Table 1, continued} Observed member stars}
\halign{
#&\quad#&\enskip#&\quad#&\enskip#&\enskip#&\quad#&\enskip#&
\enskip#&\quad#&\quad#&\quad#
&\qquad#& 
#&\quad#&\enskip#&\quad#&\enskip#&\enskip#&\quad#&\enskip#&
\enskip#&\quad#&\quad#&\quad#
\cr
\noalign{\hrule}
\noalign{\vskip0.2truecm}
{\it ID}& {\it ID$_{\rm AOP}$} & \hfil{\it V}\hfil
& \multispan3{\it \hfill RA \hfill} 
& \multispan3{\it \hfill  Dec \hfill } & \hfil $v$ \hfil & 
\hfil $\sigma_v$  \hfil & \hfil $R_{\rm TD}$ \hfil
&\hfil \quad \hfil&
{\it ID}& {\it ID$_{\rm AOP}$} & \hfil{\it V}\hfil 
& \multispan3{\it \hfill RA \hfill} 
& \multispan3{\it \hfill  Dec \hfill } & \hfil $v$ \hfil & 
\hfil $\sigma_v$  \hfil & \hfil  $R_{\rm TD}$ \hfil \cr
\noalign{\vskip0.2truecm}
308&&18.26&17&19&49.43&+57&51&00.7&-276.42&0.89&11.3&&343&&18.80&17&20&59.36&+57&47&22.1&-301.45&1.84&4.8\cr
309&21845&18.04&17&18&32.90&+57&50&56.4&-289.93&1.25&8.4&&344&&18.78&17&20&34.57&+57&47&17.9&-272.26&3.17&2.6\cr
310&&18.30&17&20&31.72&+57&50&55.2&-285.32&1.07&10.3&&346&VI-5&17.88&17&20&40.25&+57&47&07.2&-299.63&2.37&3.9\cr
311&522&18.18&17&20&13.45&+57&50&51.8&-283.24&0.92&12.2&&350&VIII-25&17.94&17&19&03.66&+57&46&42.7&-282.53&1.88&5.3\cr
312&&18.71&17&20&39.77&+57&50&50.2&-301.66&3.50&2.2&&351&V-27&18.02&17&21&39.42&+57&46&40.4&-310.59&1.32&7.7\cr
315&3237&17.58&17&20&33.59&+57&50&19.7&-277.00&2.20&3.6&&352&&18.86&17&21&00.24&+57&46&38.0&-303.39&1.73&5.5\cr
316&3210&18.11&17&20&05.43&+57&50&18.3&-288.99&1.33&7.6&&354&VII-8&17.81&17&20&09.82&+57&46&29.7&-293.85&0.82&13.4\cr
317&IV-14&17.56&17&21&30.33&+57&50&16.4&-304.75&1.07&9.5&&356&&18.44&17&19&34.43&+57&46&11.9&-286.53&1.45&6.2\cr
318&&18.89&17&19&22.60&+57&50&13.1&-300.34&2.20&4.6&&357&&18.98&17&20&18.66&+57&46&01.5&-304.12&1.62&6.0\cr
319&&18.82&17&19&32.88&+57&49&49.0&-289.04&3.27&2.4&&358&&19.01&17&22&27.17&+57&45&54.7&-294.71&2.07&4.3\cr
320&3213&18.17&17&20&11.71&+57&49&36.5&-294.10&1.36&7.8&&362&&18.90&17&22&03.68&+57&45&07.5&-287.88&1.85&5.3\cr
324&22209&17.63&17&20&21.19&+57&49&27.3&-294.40&0.77&14.8&&363&&18.82&17&20&55.60&+57&44&34.7&-304.18&2.75&2.7\cr
325&&19.34&17&19&37.77&+57&49&18.7&-292.68&5.12&1.5&&365&&19.15&17&18&56.16&+57&43&45.7&-281.69&3.71&1.8\cr
328&&18.64&17&19&20.05&+57&49&04.9&-295.53&1.30&8.3&&368&&18.41&17&18&13.07&+57&42&46.8&-286.95&2.34&3.4\cr
329&&18.52&17&20&45.48&+57&49&00.3&-282.97&1.12&10.0&&371&&18.03&17&17&46.21&+57&42&22.0&-299.06&0.93&10.6\cr
330&VIII-17&17.58&17&18&58.35&+57&48&57.7&-300.03&0.68&16.4&&375&&17.68&17&22&03.80&+57&41&28.9&-294.16&2.26&4.0\cr
331&&19.36&17&20&02.86&+57&48&57.3&-284.27&2.65&3.1&&378&&19.88&17&21&03.78&+57&39&54.6&-284.27&3.38&2.4\cr
334&VII-4&17.24&17&19&47.77&+57&48&36.7&-280.52&0.82&14.0&&383&&17.93&17&18&43.05&+57&38&50.1&-296.48&0.76&14.2\cr
337&&18.62&17&19&34.74&+57&48&05.3&-294.50&2.40&3.8&&385&&18.71&17&20&51.24&+57&38&14.6&-306.37&1.99&5.0\cr
338&&19.14&17&22&33.56&+57&48&04.0&-275.44&3.12&2.7&&386&&18.26&17&21&10.23&+57&38&02.7&-307.98&2.59&3.3\cr
339&22626&18.41&17&21&10.47&+57&47&53.5&-307.50&1.40&7.4&&400&&17.99&17&19&31.97&+57&33&35.3&-309.02&1.78&4.3\cr
342&V-26&17.41&17&21&40.41&+57&47&32.3&-293.20&0.79&13.8\cr
\noalign{\vskip0.3truecm}
\noalign{\hrule}
}
\tabnew
\endtable

\begintable*{\tabnumber}
\caption{{\bf Table 2} Probable Member stars - stars 
with velocities that indicate probable Draco membership, but with a
multi-modal or highly asymmetric cross-correlation function.  
Columns are the same as in Table 1.}
\halign{
#&\quad#&\enskip#&\quad#&\enskip#&\enskip#&\quad#&\enskip#&
\enskip#&\quad#&\quad#&\quad#
&\qquad#& 
#&\quad#&\enskip#&\quad#&\enskip#&\enskip#&\quad#&\enskip#&
\enskip#&\quad#&\quad#&\quad#
\cr
\noalign{\hrule}
\noalign{\vskip0.2truecm}
{\it ID}& {\it ID$_{\rm AOP}$} & \hfil{\it V}\hfil
& \multispan3{\it \hfill RA \hfill} 
& \multispan3{\it \hfill  Dec \hfill } & \hfil $v$ \hfil & 
\hfil $\sigma_v$  \hfil & \hfil  $R_{\rm TD}$ \hfil
&\hfil \quad \hfil&
{\it ID}& {\it ID$_{\rm AOP}$} & \hfil{\it V}\hfil 
& \multispan3{\it \hfill RA \hfill} 
& \multispan3{\it \hfill  Dec \hfill } & \hfil $v$ \hfil & 
\hfil $\sigma_v$  \hfil & \hfil  $R_{\rm TD}$ \hfil \cr
\noalign{\vskip0.2truecm}
3&&18.78&17&20&36.06&+58&23&40.8&-292.36&7.86&1.6&&273&335&18.18&17&20&06.97&+57&52&37.3&-274.88&3.88&2.0\cr
13&&18.85&17&18&06.88&+58&15&55.5&-289.49&4.75&1.1&&276&&19.10&17&18&31.14&+57&52&28.1&-302.25&4.23&1.2\cr
18&&18.69&17&21&42.56&+58&14&33.1&-276.97&4.35&2.4&&277&&19.56&17&20&45.61&+57&52&24.8&-302.09&5.58&0.8\cr
30&&19.39&17&20&24.22&+58&11&20.5&-328.94&4.82&1.6&&281&&19.51&17&20&13.23&+57&52&19.8&-291.10&4.74&1.4\cr
64&&19.09&17&20&06.08&+58&02&50.3&-277.76&4.05&1.5&&293&&18.89&17&20&44.60&+57&51&39.7&-288.47&3.49&2.3\cr
75&&18.88&17&19&57.95&+58&01&20.1&-307.67&3.89&1.6&&304&&19.74&17&21&22.25&+57&51&03.9&-284.71&4.53&0.8\cr
80&&19.62&17&21&07.79&+58&01&09.5&-313.54&4.07&0.8&&305&&18.96&17&19&18.69&+57&51&03.7&-293.41&4.19&1.9\cr
140&&19.69&17&20&55.57&+57&57&20.5&-269.48&8.27&1.4&&333&&19.44&17&20&16.57&+57&48&40.5&-288.42&3.89&1.7\cr
153&&19.64&17&21&38.81&+57&56&57.3&-285.46&4.48&1.2&&336&&19.67&17&20&59.09&+57&48&16.2&-290.32&4.08&1.6\cr
164&&19.27&17&18&45.70&+57&56&29.3&-287.29&3.45&2.1&&345&&19.14&17&19&04.39&+57&47&08.9&-294.31&3.24&2.2\cr
203&&19.64&17&20&15.90&+57&55&02.9&-303.53&5.17&1.1&&374&&19.45&17&18&50.95&+57&41&37.7&-288.88&4.98&1.2\cr
236&&19.21&17&17&56.52&+57&54&00.7&-287.88&3.09&2.4&&376&&19.50&17&17&33.69&+57&41&02.1&-296.79&3.34&1.8\cr
254&&19.64&17&19&56.29&+57&53&29.7&-296.04&4.48&1.1&&401&&19.66&17&19&55.18&+57&33&00.7&-272.36&4.30&1.1\cr
255&&19.02&17&20&41.62&+57&53&16.8&-285.73&3.64&1.9\cr
\noalign{\vskip0.3truecm}
\noalign{\hrule}
}
\tabnew
\endtable

\section{Binaries}

In this section, we first discuss the importance of corrections for
binaries and then compare the velocities of stars in our dataset with
the previous data set of AOP. We use the disagreement between the two
to constrain the number of radial velocity variables and hence Draco's
binary fraction.

\subsection{The Effects of Binaries}

When analysing kinematical data, it is important to take into account
the effect of binary stars on the velocity distribution.  Binaries
tend to inflate the apparent velocity distribution. Equally important,
the presence of a binary--induced tail increases the uncertainty of
the measured velocity dispersion, an effect that needs to be taken
into account carefully when reporting the kinematical properties of
dSphs.  Past work has shown that binary stars probably do not explain
the gross velocity dispersions of the dSphs.  Olszewski, Pryor \&
Arnamdroff (1996) used Monte Carlo simulations to demonstrate that
binary stars have a negligible effect on the velocity dispersion and
the inferred mass-to-light ratio of both Ursa Minor and Draco.
However, they also concluded that, in the decade of binary period
around one year, binary stars may be 3 to 5 times more abundant in
these two dSphs than in the solar neighbourhood.  Hargreaves, Gilmore
\& Annan (1996) used similar modeling to show that binary stars could
affect the observed velocity distribution only if the distribution of
orbital parameters were significantly different from that of solar
neighbourhood binaries.

Nevertheless, these findings are not completely general.  Although
binaries may not affect the velocity dispersion significantly, they may
nevertheless have an influence on more sophisticated forms of
dynamical model fitting.  The presence of binaries may give a
distribution that is qualitatively different from a binary-free
distribution: for example, binaries could produce a velocity tail that
is incompatible with a truncated distribution like a King model.
Empirically, we note the inclusion of binaries into a maximum
likelihood analysis of the velocity dispersion usually does not
significantly shift the most likely dispersion of a dataset; it does,
however, tend to expand the error bars, so that omitting binaries may
lead one to overestimate the statistical significance of observed
spatial variations of the dispersion.  For these reasons, we include
plausible binary distributions into our analyses.

\beginfigure*{\fignumber}
\fignam{\figstarcount}
\centerline{\psfig{figure=binary_cumulprob.ps,width=\hssize,angle=270}}
\smallskip\noindent
\caption{{\bf Figure 5.} 
Cumulative probability distribution of the absolute value of the
radial velocities of binaries drawn from two models.  The solid
(unevolved) model is taken from the Duquennoy \& Mayor (1991) solar
neighbourhood binary distribution; the dashed (evolved) model is
obtained by circularising the orbits of the unevolved model, and then
reducing angular momentum by 20 per cent, in a crude approximation of
tidal dragging caused by giant branch evolution.}
\fignew
\endfigure

For the purpose of this work, we draw velocity measurements from the
Duquennoy \& Mayor (1991, hereafter DM) binary data set.  This is a
set of binary orbital elements drawn from a sample of 165 solar--type
primary stars in the solar neighbourhood.  Of the stars in the DM
sample, 71, or 41 per cent, have companions.  Fifty-two of these 71
multiples have known binary orbital elements.  The distribution of the
period $P$ is fit by a log-normal profile, the eccentricity $e$ is fit
by Gaussian ($P<10\rm\,days$) or boxcar ($P>10\rm\,days$)
distributions, and the mass ratio $q$ is fit by a truncated Gaussian.
Using the DM orbital element distribution, we construct a cumulative
distribution of binary radial velocities and use this to build an
interpolation table of the differential binary radial velocity
distribution.  We then convolve the differential distribution with
model radial profiles to obtain the observed model profile.

Additionally, we also construct a modified DM distribution with which
we simulate giant branch binary evolution. We draw binaries from the
empirical DM distribution, discard those binaries with a periastron
$<30 R_\odot$, circularise the orbit by setting $e=0$ and making the new
circular radius equal to the original semilatus rectum, and, finally,
reduce the angular momentum by 20 per cent.  This picture of binary
evolution is justified by noting that the semilatus rectum is a
function of angular momentum, which is largely preserved during the
circularisation phase.  The 20 per cent magnitude of the subsequent
angular momentum change is at the upper range of the 10 to 20 per cent
range observed in simulations (Hurley, Tout, \& Pols 2001).  Figure 5
shows the cumulative probability distribution of radial velocities
drawn from the unevolved and evolved DM distributions.  It is evident
that binary evolution significantly increases the number of fast,
short period binaries. For instance, evolution doubles the number of
binary stars observed, at any moment, with a velocity $>10$
$\kms$.  Most of the difference between the evolved and
un-evolved distributions arises during orbit circularisation, and
substituting a 10 per cent angular momentum loss for the 20 per cent
used has little effect on the resulting binary velocity distribution.

\subsection{Draco's Binary Fraction}

Evidence on the binary fraction can be gleaned on comparison of our
dataset to previously obtained velocities in Draco.  The largest set
of Draco observations overlapping with our dataset is the 91
velocities obtained by AOP.  The 19 star data set of Hargreaves et
al. (1996) also covers the central region of Draco, but is redundant
for the purpose of our comparison, given its good agreement with the
larger dataset of AOP.  Furthermore, AOP claim that Hargreaves et
al. (1995) underestimate their velocity errors, so that any
calibration of our dataset with that of Hargeaves et al.~(1995) would
not be independent of the dataset of AOP.

The data of AOP were obtained with the KPNO 4m telescope and the HYDRA
multifibre spectrograph.  These data were obtained over a period of
nearly two years and covered a spectral range was 4720 to 5460 \AA\,
containing $\rm H\beta$ and the Mg triplet.  Velocity uncertainties
were computed both by using simulated data to rescale FXCOR errors,
and by measuring the scatter of measurements at multiple
epochs. Hence, the velocity uncertainty estimates of these data are on
a firm footing and can be used to probe the validity of our own
errors. Although we made no special attempt to re-observe the stars of
AOP, 61 of our 159 quality Draco members are also present in the previous
dataset.

\beginfigure*{\fignumber}
\fignam{\figstarcount}
\centerline{\psfig{figure=prev_compare.ps,width=\hssize,angle=270}}
\smallskip\noindent
\caption{{\bf Figure 6.} 
Difference (in units of Gaussian $\sigma$) between our measured Draco
stellar velocities and the velocities of the same 61 stars measured
previously by AOP. Although there is an excess of stars between 2 and
3$\sigma$ and the $\chi^2$ per degree of freedom is $\chi^2/N_{\rm
DOF}=1.77$, this large $\chi^2$ can be explained with binaries.
Indeed, the absence of large outliers argues against the claim of a
very large binary fraction made by AOP.}
\fignew
\endfigure

Figure 6 depicts a histogram of the
disagreement between the two datasets in units of Gaussian $\sigma$,
after subtracting a $1.22$ $\kms$ mean offset.  Although there are
no $3\sigma$ outliers, there is an apparent excess of points between 2
and 3$\sigma$.  This disagreement is reflected in the $\chi^2$ per
degree of freedom of $\chi^2/N_{\rm DOF}=1.77$ with
$p(>\chi^2)=0.00035$.

However, this disparity is readily explained by the presence of
binaries.  We ran a simulation of the disagreement between our data
set and that of AOP, contaminating both sets of data in parallel with
binaries drawn from the distribution of DM.  For this simulation, we
assumed various binary fractions and a 5 year baseline between
measurements.  In the presence of binaries, we found that 32 per cent
($1 \sigma$) of the time, $\chi^2/N_{\rm DOF}>2.4$ for unevolved
binaries with a low 20 per cent binary fraction.  For a more realistic
50 per cent binary fraction, $\chi^2/N_{\rm DOF}>7.5$ 32 per cent of
the time.  Similarly, an 80 per cent unevolved binary fraction is {\sl
ruled out} with 95 per cent certainty because the observed $\chi^2$
disparity is too small.  An evolved binary distribution, which has
a greater number of rapid orbits arising from angular momentum loss, is
even more strongly constrained: a 50 per cent evolved binary fraction
is ruled out with 0.95 certainty, and an 80 per cent evolved fraction
is ruled out at the 99.7 per cent ($3\sigma$) level.

The data are completely consistent with a binary distribution and
fraction comparable to those in the solar neighbourhood.  Unless
either we or AOP significantly {\sl over}-estimated our velocity
uncertainties, creating a very low intrinsic $\chi^2$ that would
counteract the binary-induced component of $\chi^2$, then a very large
binary fraction such as suggested by AOP is ruled out at the 2 to
3$\sigma$ level.  Even if we repeat the simulations, assuming that we
over-estimated our errors by factor of 2, we can still rule out an
unevolved binary fraction exceeding $80$ per cent and an evolved
binary fraction exceeding $50$ per cent at the 0.95 significance
level.

An immediate question that arises is why our analysis of essentially
the same data gives a different result for the binary fraction than
the original work of AOP.  AOP regarded individual $\chi^2$ outliers
as binaries, effectively considering only binaries with velocities
that disagreed between measurements at the $p=0.001$ ($3.3\sigma$)
level.  From the presence of six apparent binaries in the combined
Ursa Minor and Draco datasets, AOP extrapolated a binary fraction
several times larger than that of the solar neighbourhood, with large
uncertainties.  However, these $\chi^2$ outliers could have resulted
from a mild misestimation of the velocity errors for these stars.  Our
upper limit on the binary fraction, however, is based on the 
absence of outliers and is sensitive to an {\sl over}-estimation of
velocity errors. For these reasons, we believe that our analysis is more
likely to be robust against error misestimation or data corruption,
and that there is no evidence supporting a large binary fraction.

\beginfigure*{\fignumber}
\fignam{\figvdisp}
\centerline{\psfig{figure=vdisp_fb_0.0_no_rotsub.ps,width=6.5cm,angle=270}
            \hskip 1.5cm
            \psfig{figure=vdisp_fb_0.0.ps,width=6.5cm,angle=270}}
\vskip 0.5cm
\centerline{\psfig{figure=vdisp_fb_0.4_no_rotsub.ps,width=6.5cm,angle=270}
            \hskip 1.5cm
            \psfig{figure=vdisp_fb_0.4.ps,width=6.5cm,angle=270}}
\vskip 0.5cm
\centerline{\psfig{figure=vdisp_fb_0.8_no_rotsub.ps,width=6.5cm,angle=270}
            \hskip 1.5cm
            \psfig{figure=vdisp_fb_0.8.ps,width=6.5cm,angle=270}}
\smallskip\noindent
\caption{{\bf Figure \figvdisp.} 
Variation of line of sight velocity dispersion with projected radius;
Filled circle is the most likely velocity dispersion in each radial
bin and the inner and outer error ticks represent 68 and 95 per cent
(1 and $2\sigma$) confidence limits respectively. Figures in the left
column contain observed velocities; figures in the right column
contain velocities with the solid-body rotational component
subtracted.  From top to bottom, the figures assume an increasing
binary fraction $f_b$.  Binaries are taken from the empirical solar
neighbourhood distribution of Duquennoy and Mayor (1991) and are then
evolved through circularisation and angular momentum loss, as
described in the text.  The smooth curves are drawn from the
$\alpha,\gamma$ models of Wilkinson et al. (2001), and the two letter
codes denote the model type -- {\it t:} tangentially anisotropic
($\gamma=-6.1$, or $\nu=-0.6$); {\it i:} isotropic ($\gamma=\nu=0$); 
{\it r:} radially
anisotropic ($\gamma=1.5$, or $\nu=0.6$) -- 
{\it h:} harmonic halo ($\alpha=-2$);
{\it f:} flat rotation curve halo ($\alpha=0$); {\it m:} mass follows
light ($\alpha=1$).  All models are normalised to have a central
velocity dispersion of 8.5 $\kms$, in agreement with the
observed value as well as the published value of AOP.}
\fignew
\endfigure

\section{Observed Kinematics}

\subsection{Radial Variation of the Velocity Dispersion}

Although nearly 100 velocities have been measured in Draco prior to
this work, they have revealed little about the radial variation of the
velocity dispersion.  As previous spectroscopic targets were selected
using relatively imprecise photographic plate photometry, Galactic
foreground contamination limited velocity measurements to the dense
centre of the dSph, where dSph member yield is high.  Hargreaves et
al. (1996) found that the velocity dispersion appears to be flat or to
rise with radius, and is inconsistent with the best--fit King model
profile at the $2\sigma$ level; however their data had only three
points outside the core radius.  The data of this paper, in contrast,
contain 33 points outside $10^\prime$ (approximately the King core
radius), and include several stars very close to the King tidal radius,
where the velocity should drop to zero under the King model
hypothesis.

When computing the velocity dispersion as a function of radius, we use
a maximum likelihood technique which assumes that the velocity
uncertainties are Gaussian and that the velocity distribution in any
radial bin is a Gaussian centred on the mean velocity $\bar v$ of the
entire ensemble.  Furthermore, we assume that the stars are drawn from
a binary distribution with a radial velocity probability distribution
$P_b(v)$.  The observed velocity distribution is then the assumed
Gaussian profile convolved with observational errors and $P_b(v)$.

The probability distribution of the true velocity dispersion
$\sigma_v$ of the bin is then given by
$$
\eqalign{
P(\left<v^2\right>)\propto \prod_i \int_{-\infty}^{+\infty} &
{ \exp\left({-{1\over2} {(v_i-v^\prime-\bar v)^2 \over \sigma_i^2+\left<v^2\right>}}\right) \over \sqrt{2\pi \left( \sigma_i^2+\left<v^2\right> \right)}} \cr
 & \times \left[(1-f)\delta(v^\prime)+f\,P_b(v^\prime) \right] \,dv^\prime
}
\eqno(1)
$$
where $v_i$ and $\sigma_i$ are the $i^{\rm th}$ velocity in the bin
and its uncertainty, $f$ is the binary fraction, $\delta$ is the usual
Dirac delta function, and $\bar v$ is the mean velocity of the entire
ensemble.  In practice, we speed the computations by using Fourier
convolutions to create a library of $P(\left<v^2\right>)$
parameterised by $\sigma^2+\left<v^2\right>$.  Numerical integration
of the tabulated $P(\left<v^2\right>)$ then gives the uncertainties of
the velocity dispersion.

Figure \figvdisp\ shows a set of plots of velocity dispersions,
assuming various binary fractions.  As pointed out by Oh, Aarseth \&
Lin (1995), tidal disruption can have the appearance of rotation,
because the first order expansion of the tidal field produces a
stretching force that, in projection, causes motion similar in
appearance to rotation.  Therefore, in the bottom row of Figure
\figvdisp, we also plot the velocity dispersion for models in which
rotation on the sky has been subtracted; we accomplish this
subtraction by transforming each velocity $v_i$ to $v_i^\prime =
v_i-(\xi_x x_i + \xi_y y_i)$, where the vector $(\xi_x,\xi_y)$ is
computed by requiring the projected angular momentum $\bf L$ to equal
zero according to
$${\bf L} = 
\left(L_x, L_y\right)=
\left(\sum_i v_i^\prime y_i ,  \ 
-\sum_i v_i^\prime x_i\right) = (0,0) \,.
\eqno(2)
$$
The velocities produced by this rotation subtraction are not
necessarily dynamically meaningful because we are also subtracting
any genuine rotational component as well as any tidal component.
However, the purpose of examining rotation subtracted data is to
demonstrate that the observed variation of the velocity dispersion
with radius cannot be attributed to tidal effects.  Indeed, after
rotation subtraction, we will systematically under-estimate the
velocity dispersion at large radii because the Gaussian dispersion
model to which the data are fit do not include the effects of rotation
subtraction on a non-rotating distribution.

Furthermore, Draco will appear to have a perspective induced apparent
rotation caused by the variation of the line-of-sight angle across its
face (Fest, Thackeray, \& Wesselink 1961).  The solar motion component
of this rotation is 1 $\kms$ $30^\prime$ from the centre.  The
component from Draco's motion is poorly constrained because Draco's
proper motion is uncertain.  Monte-Carlo orbital simulations of
Draco's orbit based on its proper motion show that Draco's most
probable line-of-sight angle is $70^\circ$, implying a perspective
rotation of 2.5 $\kms$ across $30^\prime$.  However, this rotation
increases to 5 $\kms$ if the line-of-sight angle is $80^\circ$.
Analysing Draco's velocities with all rotation subtracted effectively
demonstrates that these uncertainties do not affect our dynamical
conclusions.

From Figure \figvdisp, it is apparent that the velocity rises from the
central value of 8.5 $\kms$.  In the most plausible case of no
rotation subtraction and a 40 per cent evolved binary fraction, the
last three bins have a velocity dispersion larger than the central
value with $1\sigma$, $2\sigma$, and $1\sigma$ statistical strengths,
with a combined probability $<10^{-3}$.  Only when an implausibly
large 80 per cent evolved binary fraction is combined with rotation
subtraction does this conclusion come under question; however, this
large binary fraction is ruled out by the good agreement of our
velocities with 61 prior velocities of AOP.

\beginfigure*{\fignumber}
\fignam{\figvdispgoodstars}
\centerline{
 \psfig{figure=vdisp_fb_0.4_TDR_GT_5_no_rotsub.ps,width=6.5cm,angle=270}
 \hskip 1.5cm
 \psfig{figure=vdisp_fb_0.4_TDR_GT_7_no_rotsub.ps,width=6.5cm,angle=270}}
\smallskip\noindent
\caption{{\bf Figure \figvdispgoodstars.} Variation of line-of-sight
velocity dispersion for good quality spectra with a comparatively large
Tonry \& Davis (1979) $R_{\rm TD}$ value.
Left:  $R_{\rm TD}>5$; Right:  $R_{\rm TD}>7$. }
\fignew
\endfigure

In Figure \figvdispgoodstars, we plot the variation of the velocity
dispersion with projected radius for a subsample of stars with
high-quality spectra (Tonry \& Davis $R_{\rm TD}>5$ and 7).  The
$R_{\rm TD}>5$ sample contains 95 stars, and the $R_{\rm TD}>7$ sample
contains 73 stars.  In neither case does the exclusion of low-quality
spectra affect the finding of a rising velocity dispersion.

In the $R_{\rm TD}>7$ panel of Figure \figvdispgoodstars, the velocity
dispersion falls to 4 $\kms$ in first bin, which extends from the
centre to $R=4^\prime$.  This effect is not the result of sample size
effects, because there are 16 stars in this bin, and because our
dispersion estimator is valid for small samples.  A bootstrap
resampling of 16 stars from the 23 stars in the left $R_{\rm TD}>5$
panel of Figure
\figvdispgoodstars\ shows that the observed dispersion drop 
occurs by chance 3 per cent of the time; if the binary fraction is
doubled from the 40 per cent fraction assumed, this probability rises
to 18 per cent.  Given that there are six radial bins in which such a
drop could have been observed, the drop is probably a statistical
artefact, and is not the result of the difference in spectrum
quality between the two panels.

\subsection{Direction and Significance of Rotation}

When subtracting solid-body rotation from our data set, we found that
Draco is rotating at 6 $\kms$ at $30^\prime$ about an axis with a
position angle (PA) of $62^\circ$ east from north with the northwest
side of the galaxy receding.  Applying a Monte-Carlo analysis of the
proper motion data of Scholz \& Irwin (1994), we find that the PA of 
Draco's proper motion is ${29^\circ}^{+17}_{-19}$ in the frame
moving with the sun.   Thus the perspective induced rotation axis
should be have a PA of ${119^\circ}^{+17}_{-19}$.  This is in
$57^\circ$ or $3 \sigma$ disagreement with the observed
rotation axis.

Corrected for solar motion, Draco's proper motion vector has a PA of
${42^\circ}^{+23}_{-23}$.  This is in $2\sigma$ disagreement with the
PA of the major axis of Draco at ${88^\circ}^{+3}_{-3}$ (Odenkirechen
et al. 2001), arguing against tidal stretching. Additionally, the axis
of tidally induced rotation should have a PA of
${132^\circ}^{+23}_{-23}$ because it is expected to be perpendicular
to the motion (eg. Oh, Lin, \& Aarseth 1995); this is in $70^\circ$ or
$3\sigma$ disagreement with the observed rotation axis angle of
$62^\circ$.

Additionally, we performed a Monte Carlo simulation to test the
significance of the rotation observed in our Draco data, modelling it
is as drawn from a 10 $\kms$ non-rotating Gaussian distribution. We
found that 5 per cent of the time, this non-rotating distribution had
$\bf L$ larger than that observed in the actual data.  We also
performed a more realistic simulation with stars drawn from a
non-rotating Gaussian possessing the observed radial dispersion
variation, and included binary contamination with a 40 per cent
unevolved binary population.  For this simulation, the probability of
randomly seeing an equal or larger rotation is 14 per cent.

In summary, the rotation we observe is statistically weak,
inconsistent with tides, and inconsistent with the perspective
effect.  Furthermore, the elongation of Draco is 
misaligned with Draco's proper motion at the $1.7\sigma$ level.
Thus there is no evidence of tidal effects in our Draco
data, or in the proper motion data of Scholz \& Irwin (1994).
This is in agreement with the absence of tidal truncation
noted in  SDSS Draco data (Odenkirchen et al. 2001).

The direction of our observed rotation is inconsistent with the
finding of AOP, who found rotation about an axis with a position angle
of $15^\circ$ at the $1\sigma$ significance level.  However,
Hargreaves \etal\ (1996) also found $2\sigma$ rotation about the major
axis though from a centrally concentrated sample of 17 stars.  
None of these findings are statistically significant enough
to constitute a genuine detection of rotation.

\beginfigure*{\fignumber}
\fignam{\figfit}
\centerline{\hskip 1cm \hfil 
            \psfig{figure=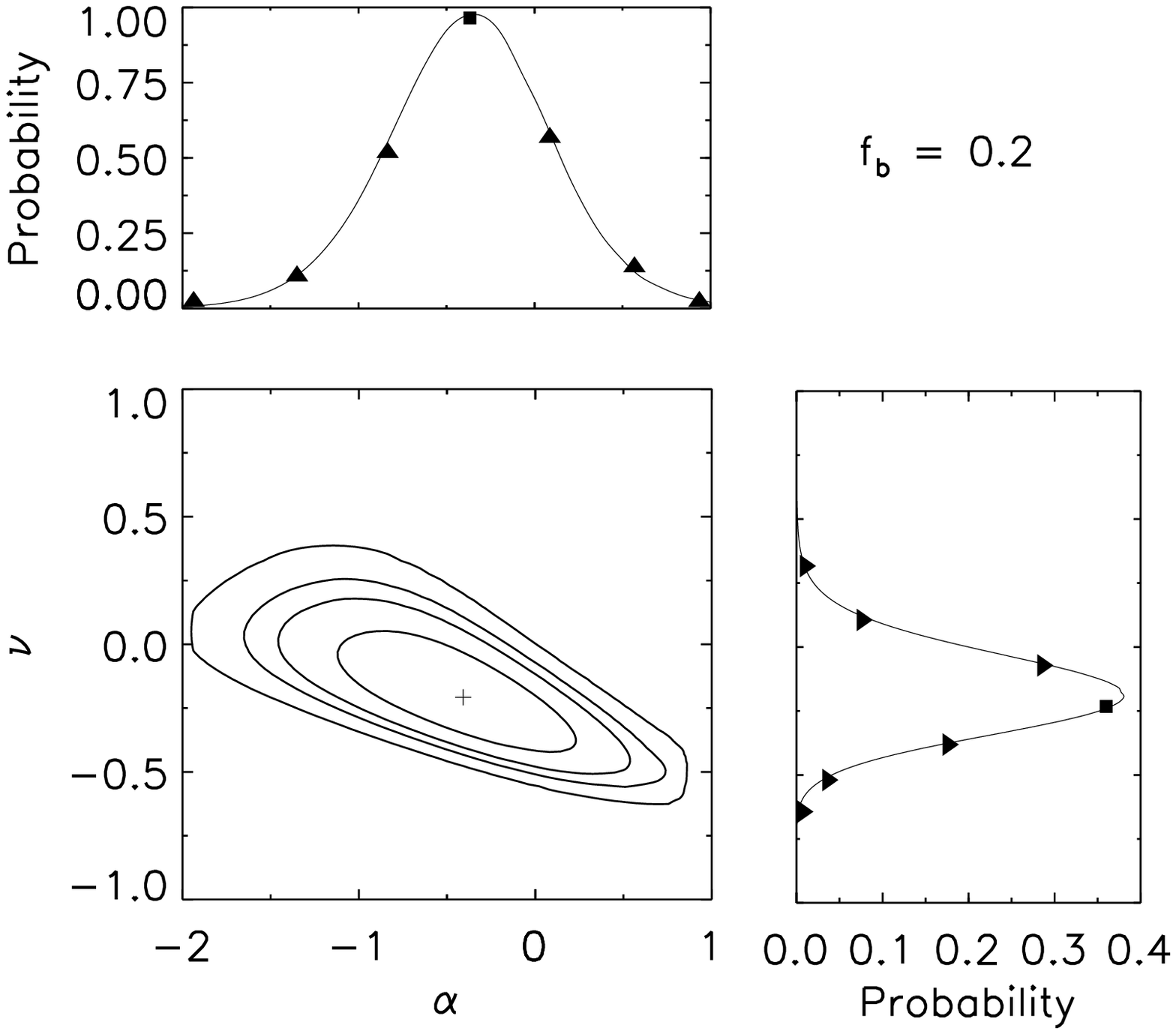,width=6.5cm,angle=0}
	    \hskip 2cm
            \psfig{figure=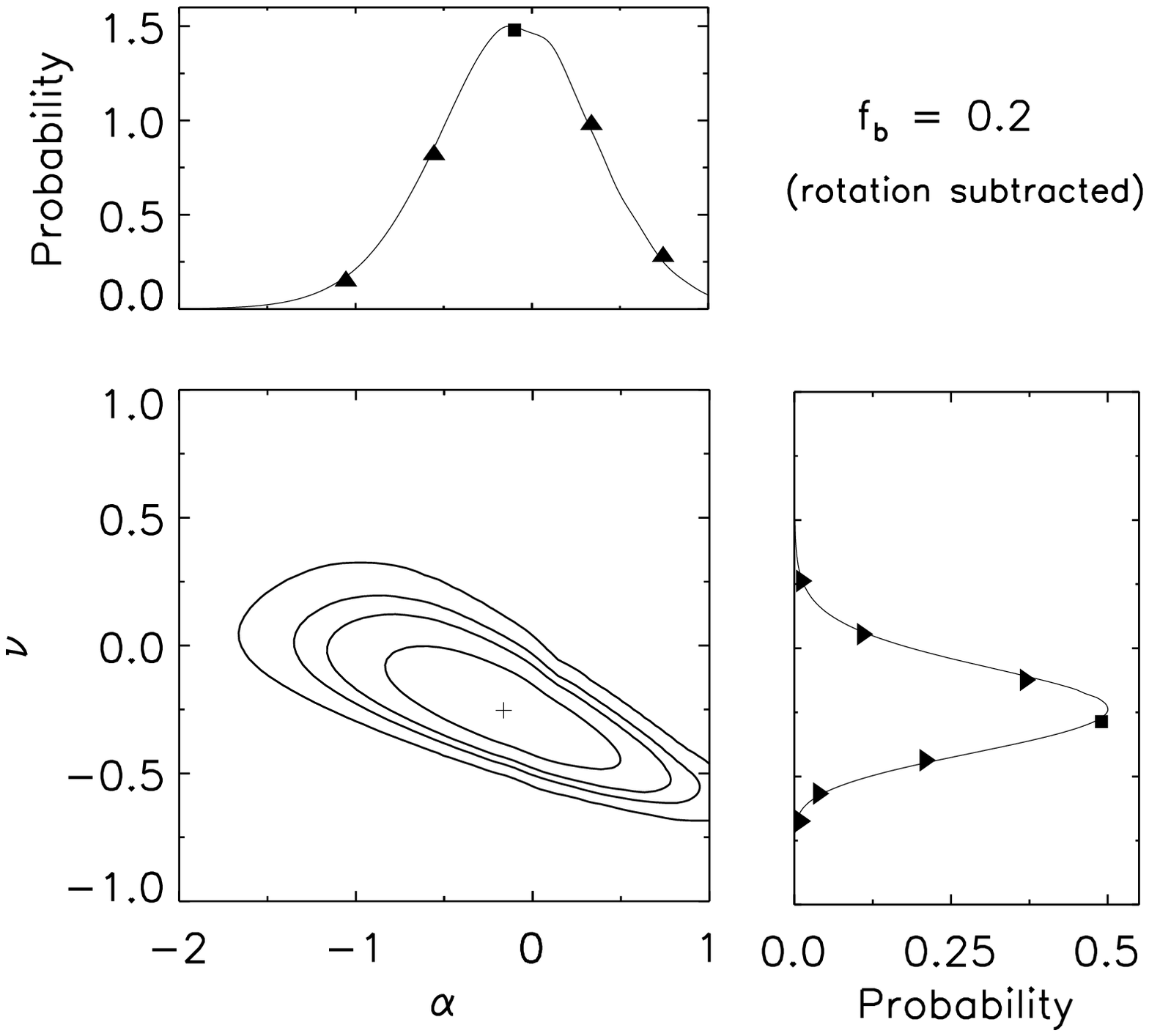,width=6.5cm,angle=0}
	    \hfil}
\vskip 1.0cm
\centerline{\hskip 1cm  \hfil 
            \psfig{figure=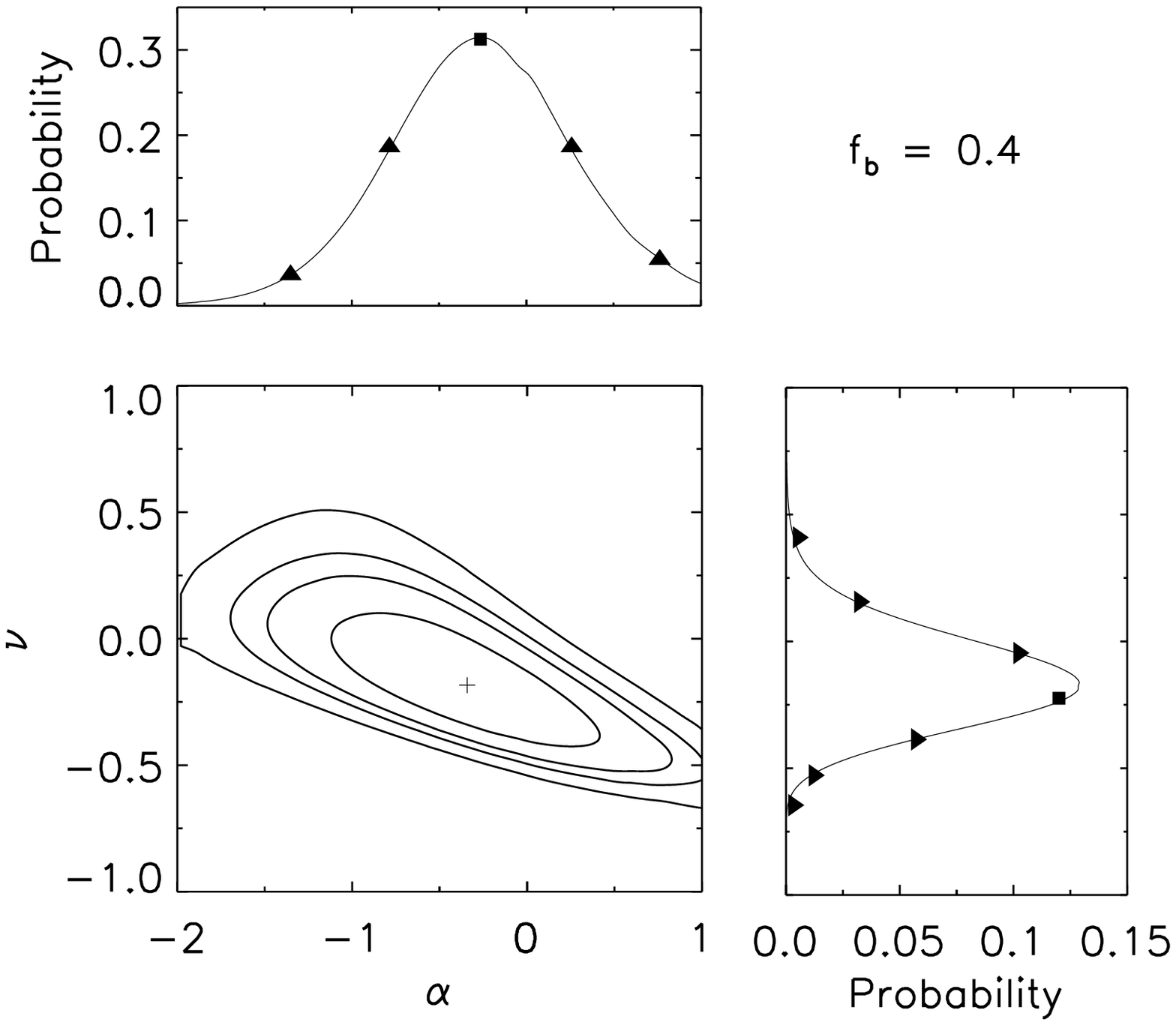,width=6.5cm,angle=0}
	    \hskip 2cm
            \psfig{figure=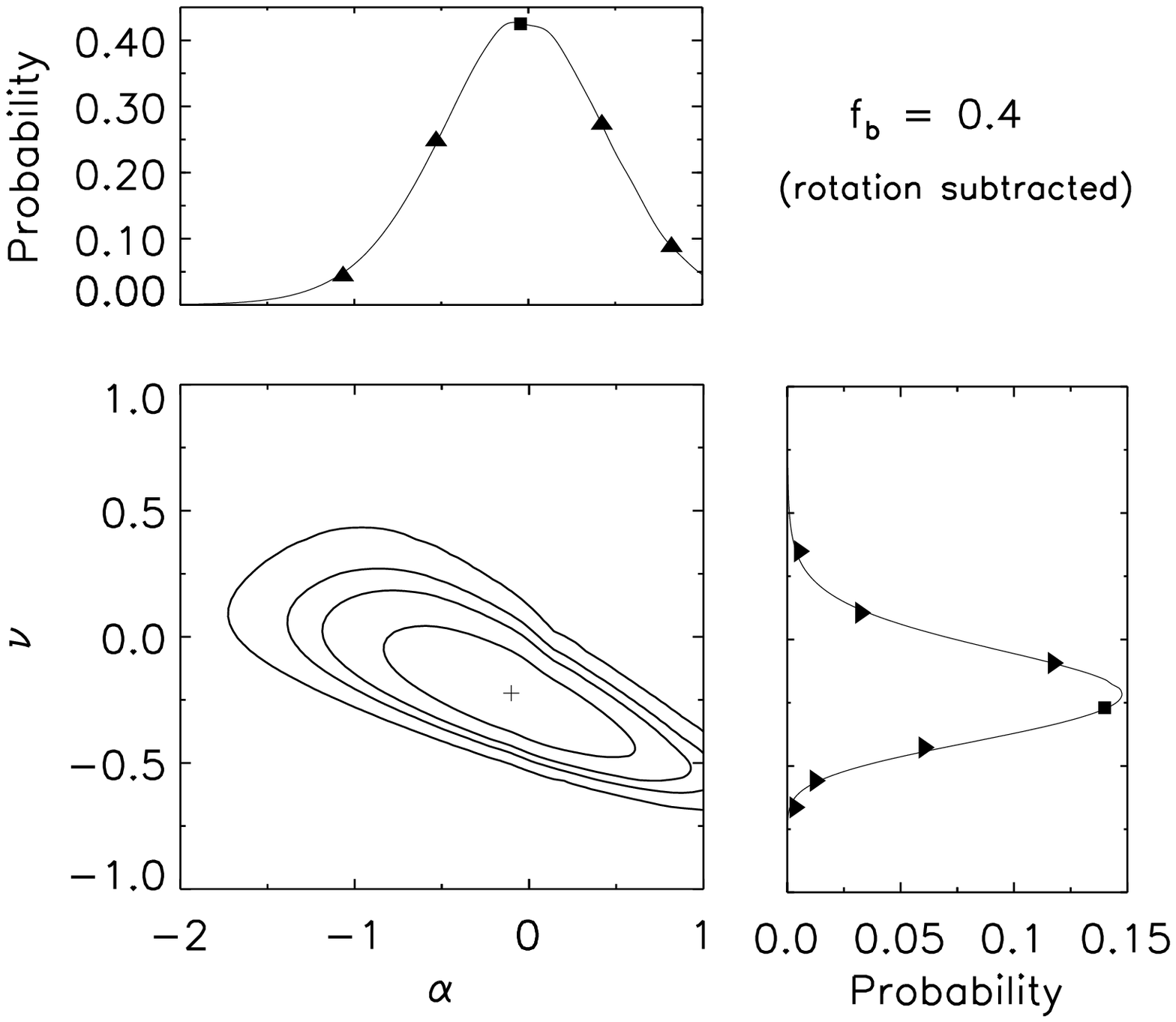,width=6.5cm,angle=0}
	    \hfil}
\vskip 1.0cm
\centerline{\hskip 1cm  \hfil 
            \psfig{figure=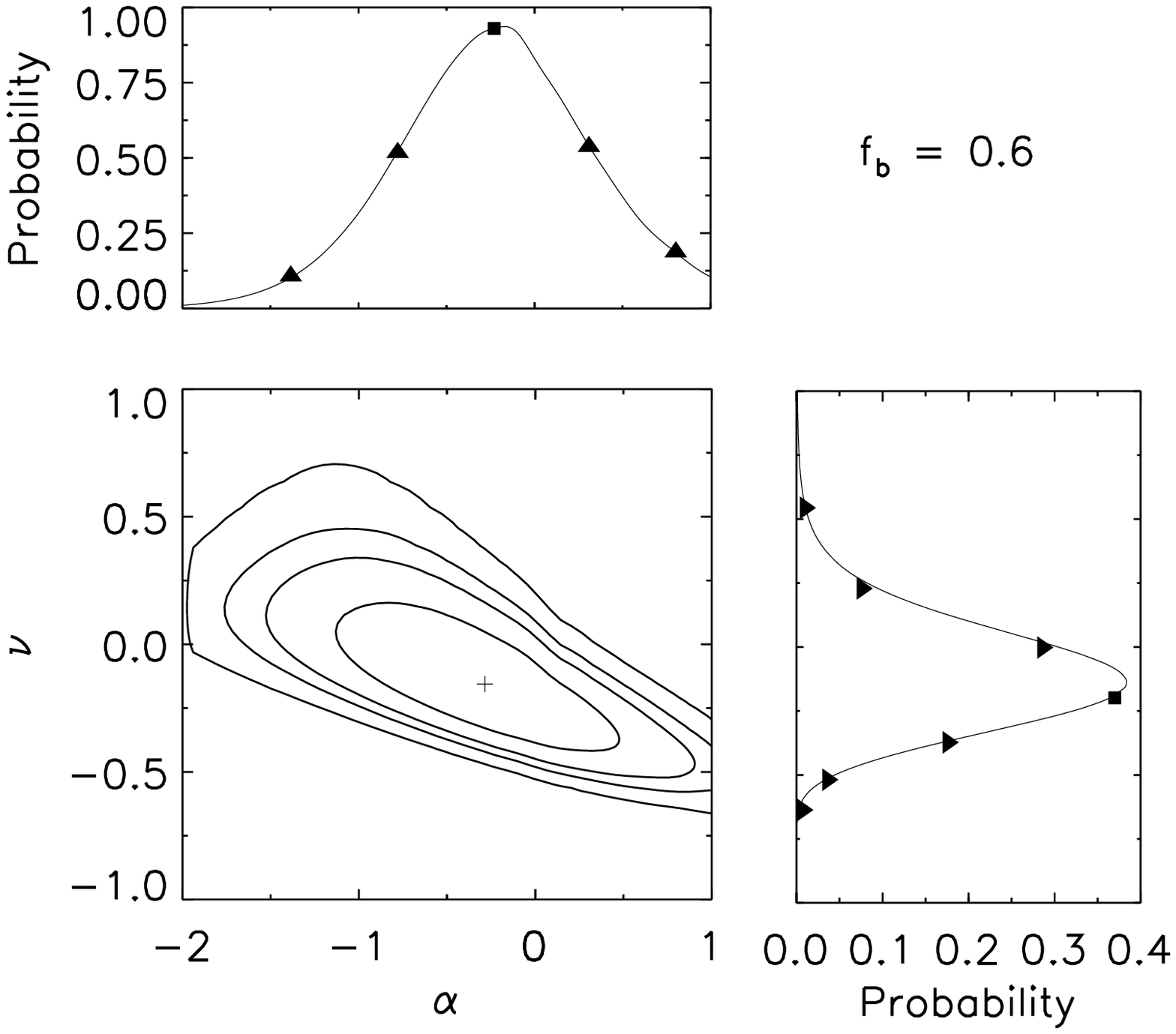,width=6.5cm,angle=0}
	    \hskip 2cm
            \psfig{figure=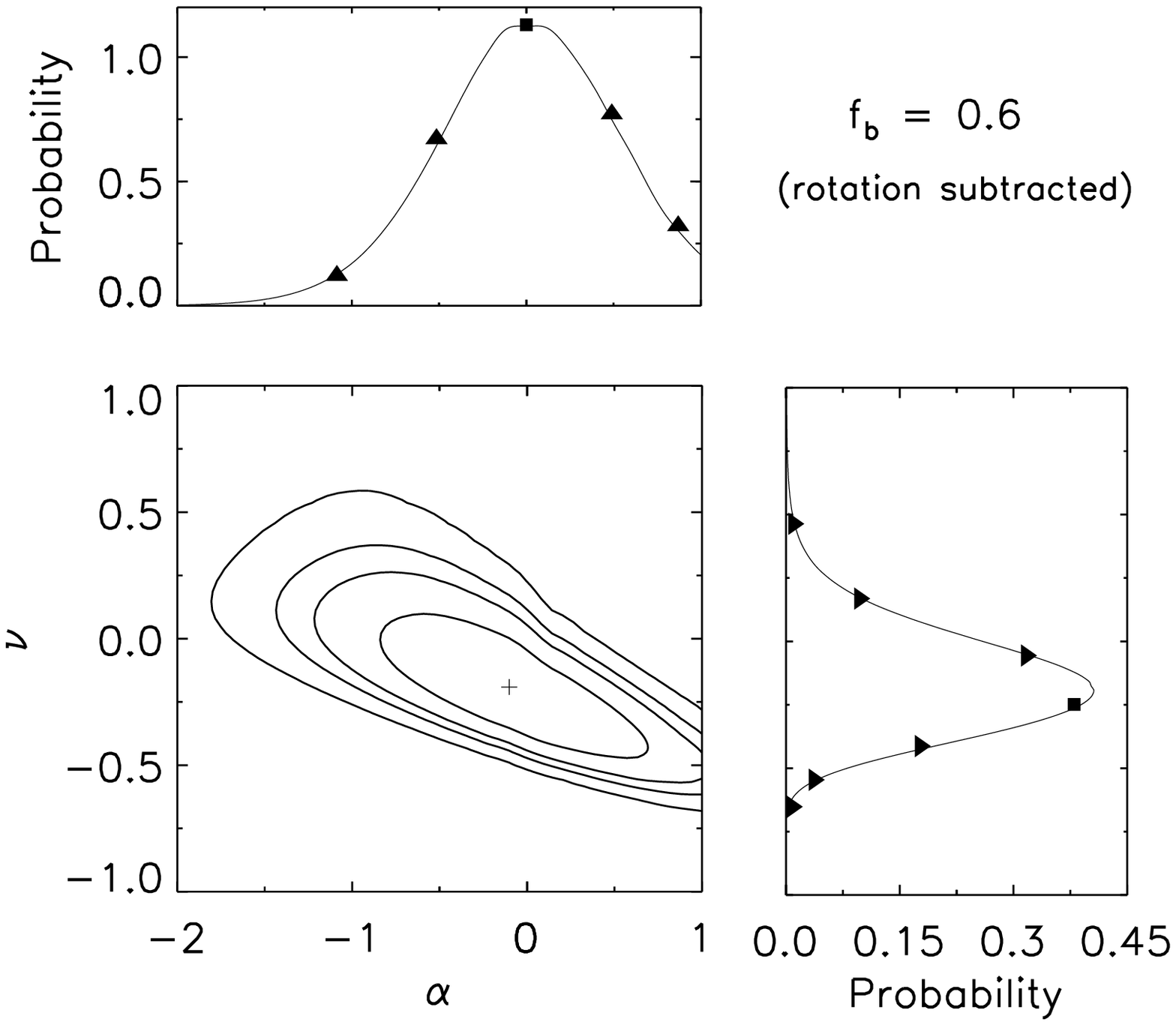,width=6.5cm,angle=0}
	    \hfil}
\smallskip\noindent
\caption{{\bf Figure \figfit.} 
Likelihood contours of the fit of our Draco data to the two-parameter
$\alpha,\gamma$ models of Wilkinson et al. (2001);
$\nu=-\log_{10}\left[\left(2-\gamma\right)/2\right]$ is used as a
symmetrical surrogate for the anisotropy parameter $\gamma$.  The
contours are at enclosed two-dimensional $\chi^2$ probabilities of
0.68, 0.90, 0.95, and 0.997.  The top and right panels of each plot
represent the probability distributions of $\alpha$ and $\nu$,
respectively; the median of each distribution is represented by a
square, and the triangles are at 0.68, 0.95, and 0.997 enclosed
probabilities.  $f_b$ is the binary fraction used, assuming evolved
binaries, as described in the text.  The left column contains analyses
of the data as observed; the right column uses data from which the
rotational component of motion has been subtracted.}
\fignew
\endfigure

\subsection{Model Fitting}

In Paper I, we introduced a set of two parameter dynamical models for
spherical stellar systems.  These models are designed to probe the
principal observational degeneracy -- namely, the trade-off between
halo mass and orbital anisotropy.  For example, a dSph like Draco,
with large velocities at extended projected radii, may have either a
large dark matter halo or may have orbits that are tangentially
anisotropic.  In the latter case, the large apparent velocity
dispersion arises from from the fact that, at large radii, the radial
velocities observed are drawn in large part from circular orbits
viewed edge-on.

The two parameters of the models are $\alpha$, which determines the
dark matter content and $\gamma$, which determines the velocity
anisotropy.  An $\alpha=1$ model corresponds to a mass-follows-light
dSph; $\alpha=0$ is a dSph with a flat rotation curve; and $\alpha=-2$
is the case of a dSph immersed in the centre of a uniform density
(harmonic) halo.  When $\gamma=2$, the orbits are radial at large
radii. When $\gamma=0$, the velocity distribution is isotropic (as is
a King model). When $\gamma=-\infty$, the orbits are all circular at
large radii.  The light distribution of these models is independent of
the potential and is always represented by a Plummer (1911) model.  As
shown in Paper I, a Plummer model with a core radius $r_0=9.7^\prime$
is a good fit to Draco's stellar surface density -- as good as the
King models that are customarily used.

Our method of fitting the data is identical to the method used for the
synthetic datasets of Paper I.  In summary, a line profile is derived
from the model by numerically integrating out the tangential
velocities.  This line profile is tabulated in a multi-dimensional
spline table. This allows rapid generation of the line-of-sight velocity
($v_{\rm los}$) line profile
$L(v_{\rm los},R;\alpha,\gamma)$ for any value of projected radius $R$
and $\alpha,\gamma$.  The line profile is then convolved with a binary
distribution and Gaussian velocity measurement uncertainty profile
$\sigma$ to give $\tilde L(v_{\rm los},R, \sigma;\alpha,\gamma)$. 
The probability of observing an ensemble of points at $R_i$ with
velocity and velocity uncertainty $v_i,\sigma_i$ is 
$$
P(\lbrace R_i,v_i,\sigma_i\rbrace | \alpha,\gamma ) 
= \prod_i \tilde L(v_i,R_i,\sigma_i;\alpha,\gamma)
\eqno(2)
$$
Finally, $P(\lbrace R_i,v_i,\sigma_i\rbrace | \alpha,\gamma )$
is tabulated over a grid of $\alpha,\gamma$, and the
resulting likelihood contours and their projections onto
the $\alpha,\gamma$ axes are used to constrain the best-fit
$\alpha,\gamma$.

\beginfigure{\fignumber}
\fignam{\figfittwo}
 \psfig{figure=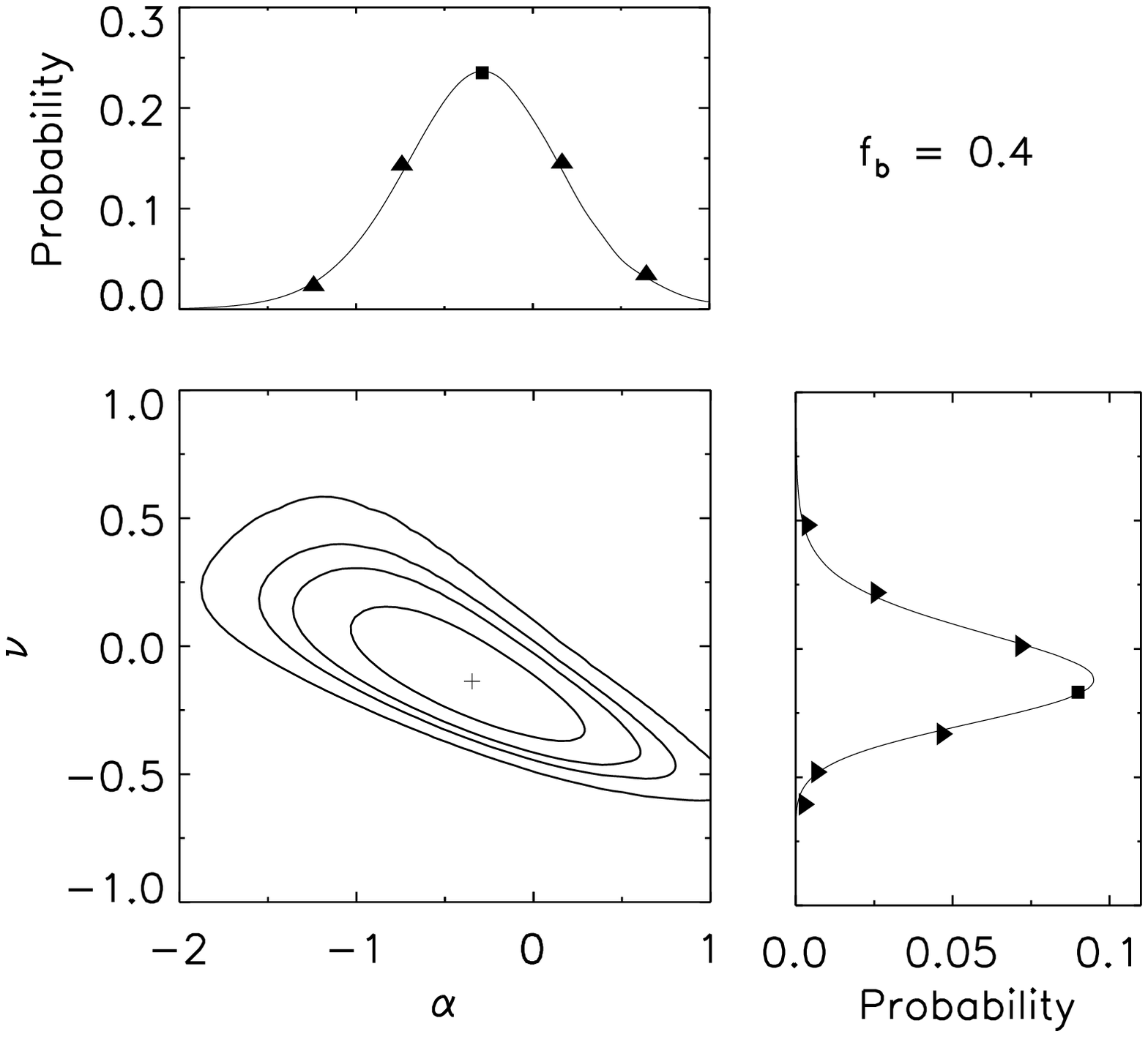,width=6.5cm,angle=0}
 \vskip 1cm
 \psfig{figure=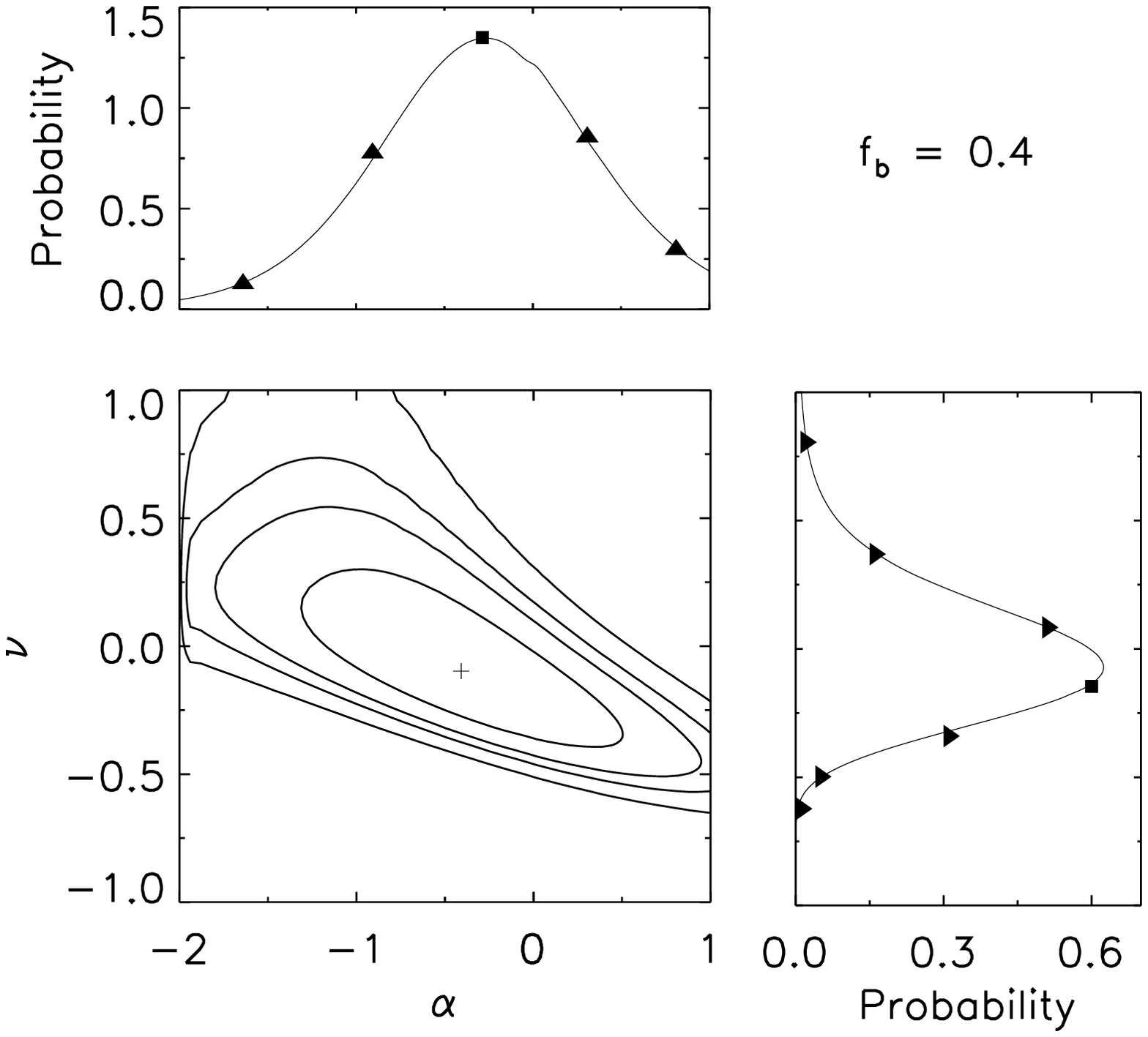,width=6.5cm,angle=0}
 \vskip 1cm	
 \psfig{figure=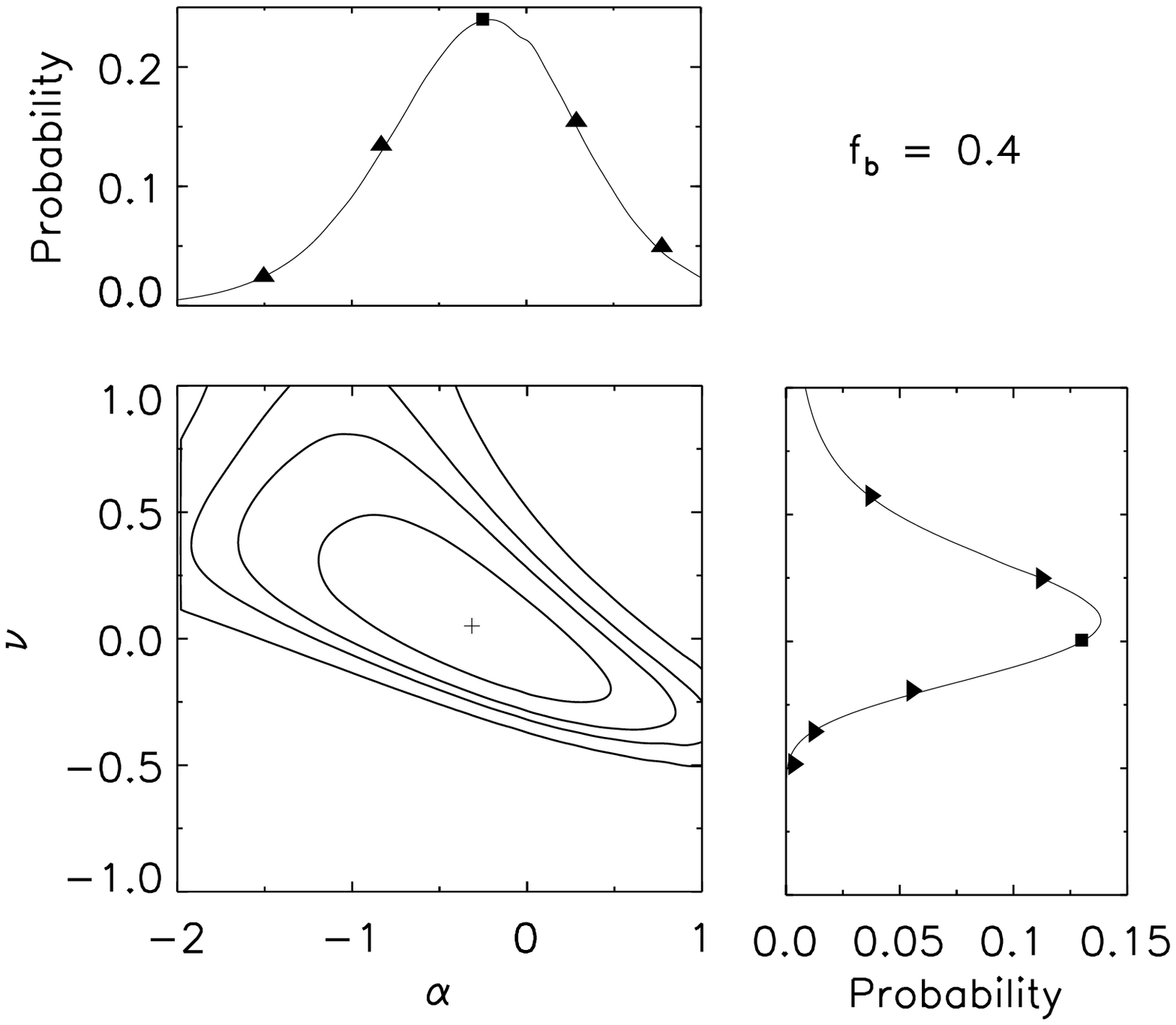,width=6.5cm,angle=0}	
\smallskip\noindent
\caption{{\bf Figure \figfittwo.} The model fits of Figure~\figfit\
repeated with the introduction of deliberate errors.  Top: The Plummer
core radius of the light has been increased by 20 per cent over the
best-fit value.  Centre: The velocity uncertainties have been
increased by a factor of $\sim 2$ by setting all velocity
uncertainties less than 2 $\rm km\,s$${}^{-1}$ equal to 4 $\rm km\,s$${}^{-1}$.
Bottom: The central velocity dispersion of the model has been set to
10 $\rm km\,s$${}^{-1}$, rather than the observed value of 8.5 
$\rm km\,s$${}^{-1}$.
}
\fignew
\endfigure

Figure~\figfit\ shows the result of fitting our Draco velocity data
with the two-parameter dynamical model for various combinations of
binary fraction and rotation-subtraction.  The fits are presented
as likelihood contours in the two dimensional parameter space of
the model.  
Here, we have symmetrised
the $\gamma$ parameter describing the anisotropy by defining a
new anisotropy parameter
$\nu=-\log_{10}\left[\left(2-\gamma\right)/2\right]$.  The value
$\nu=1$ then corresponds to a system in which
$\left<v_r^2\right>/\left<v_\theta^2\right>\rightarrow10$ in the limit
of large radii; $\nu=0$ is isotropic; and $\nu=-1$ corresponds to a
system in which $\left<v_r^2\right>/\left<v_\theta^2\right>\rightarrow
1/10$.  The left panels use the observed data, and the right panels use
rotation subtracted data.  In all instances, including the extreme
case of rotation-subtracted data and a 0.6 evolved binary fraction, a
mass-follows-light model ($\alpha =1$) is ruled out at the 2 to $3\sigma$
level.  Likewise, in all cases, an extended halo with a harmonic core
($\alpha = -2$) is ruled out at the 2 to $3\sigma$ level.  

Assuming a 40 per cent binary fraction and omitting rotation
subtraction produces the best fit values $\alpha=-0.34$, and
$\nu=-0.18$ in two dimensions.  For the projected (one-dimensional)
distributions, the median values are $\alpha=-0.26$ with a $1\sigma$
range -0.78 to 0.25, and $\nu=-0.22$ with with a $1\sigma$ range -0.39
to -0.05.  These values correspond to a dSph with a weakly
tangentially anisotropic distribution of velocities. At large radii,
the velocity dispersions have the property
$\left<v_r^2\right>/\left<v_\theta^2\right>\rightarrow 0.56$.  The
dark matter is distributed in a nearly isothermal halo. More exactly,
the rotation curve of the halo is very slowly rising like $r^{0.17}$
at large radii.  Extremely anisotropic velocity distributions, as well
as halo models markedly deviating from isothermal, are ruled out.

Figure~\figfittwo\ checks the robustness of our maximum likelihood
solution against possible errors. The dataset is now analysed with
an assumed luminous Plummer core radius  20 per
cent larger than the best-fit value (top), or assuming that the velocity
uncertainties have been underestimated by a factor of $\sim 2$
(centre), or with an assumed central velocity dispersion of 10 $\kms$
(bottom).  In all cases, the maximum likelihood solution
barely changes.  The fact that a large mis-estimation of the 
Plummer core radius does not affect the result strongly suggests
that the exact form of the parameterised stellar density is
not important.

\beginfigure{\fignumber}
\fignam{\figfitgood}
 \psfig{figure=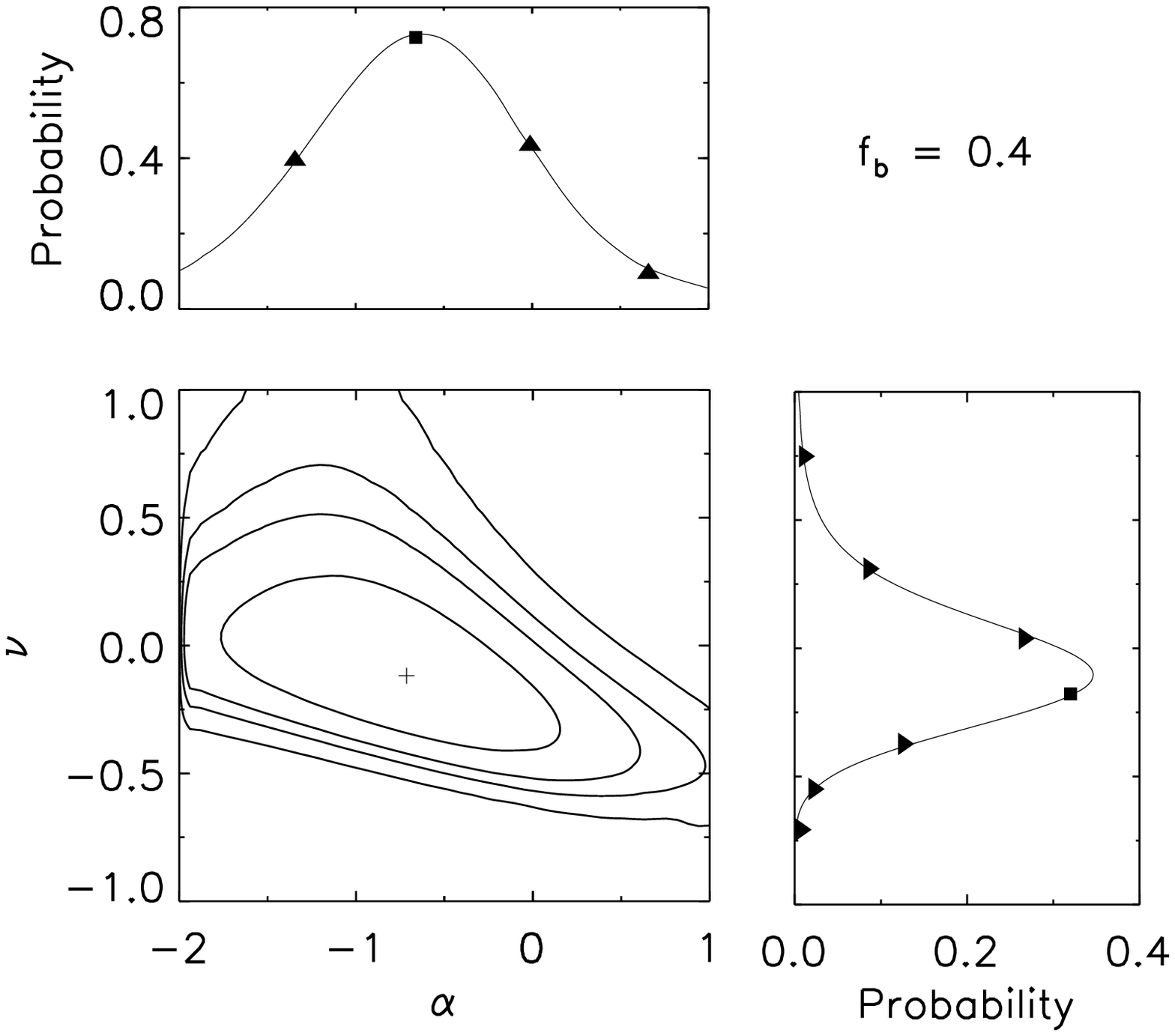,width=6.5cm,angle=0}
 \vskip 1cm
 \psfig{figure=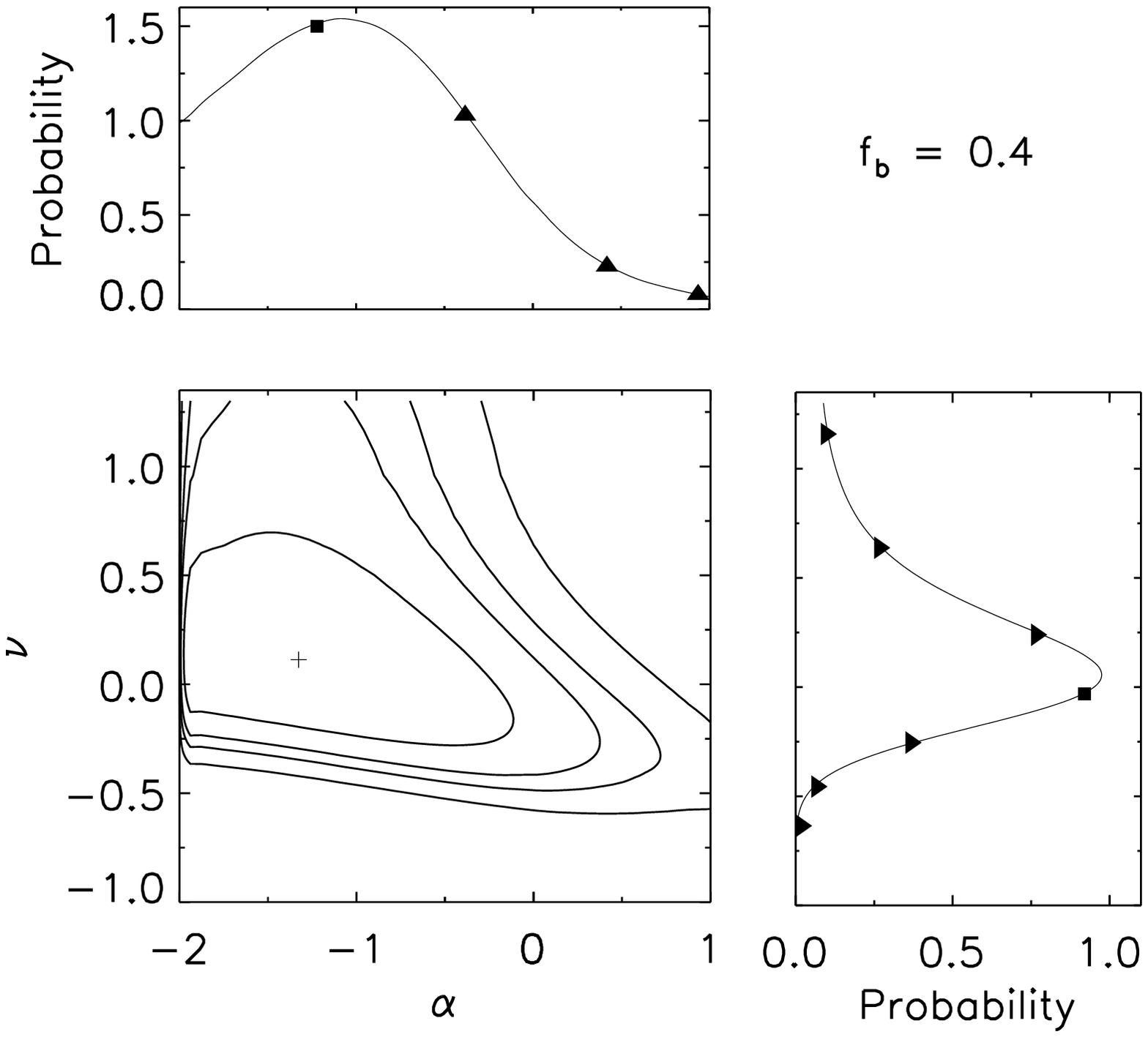,width=6.5cm,angle=0}
\smallskip\noindent
\caption{{\bf Figure \figfitgood.} The model fits of Figure~\figfit\
repeated using stars with good quality spectra only.  Top:
spectra with Tonry \& Davis $R_{\rm TD}>5$; Bottom:
spectra with Tonry \& Davis $R_{\rm TD}>7$.}
\fignew
\endfigure

Finally, Figure ~\figfitgood\ shows the results of performing the model
fit with good quality spectra only.  Although the contours are
considerably broader than before, the $\alpha=1$ mass-follows-light
model is ruled out at a comparable statistical signficance.
The most likely model has a somewhat more massive halo and
is less isotropic than when the fit was performed with
the entire sample.  However, the two fits are consistent
well within the $1\sigma$ level.  Hence it is quite unlikely
that our finding of a largely isotropic distribution function
inside an extended halo is the result of velocity errors
in low-quality spectra.

From the best--fit $\alpha,\nu$ values of Figure \figfit, it is
possible to constrain Draco's mass-to-light ratio, at least within the
region in which data exist.  For a typical $\alpha=-0.3, \nu=-0.25$
best-fit model, the mass within three core radii is
${8^{+3}_{-2}}\times10^7 M_\odot$.  The mean mass-to-light ratio within
3 core radii is $(330\pm125) M_\odot/L_\odot$, assuming a
luminosity of $(2.4\pm 0.5)\times 10^5 L_\odot$ (Odenkirchen \etal\
2001).  Although our mass superficially appears consistent with AOP
(1995) and Hargreaves \etal\, the mass we measure is actually the
inner part of a halo, rather than the total mass of a
mass-follows-light distribution.

\section{Conclusions}

We have carried out a radial velocity survey of the Draco dwarf
spheroidal (dSph) galaxy using the AF2/WYFFOS instrument at the
William Herschel Telescope. A total of 284 stars were targeted. Of
these, 203 stars had usable velocities, 186 stars turned out to be
Draco members and 159 stars had velocities of sufficient quality for
our subsequent dynamics analyses.  As our initial target list was
based on accurate CCD data, rather than photographic plate photometry,
we achieved a clean separation between Draco's giant branch population
and the Galactic foreground out to the tidal radius.  Consequently,
our dataset contains many more stars at large radii than previous
studies.  We calibrated the measurement uncertainties of our dataset
by subdividing the bright stars into two velocity measurements, and
using the discrepancies between the measurements to rescale the
nominal IRAF Tonry-Davis $R$ value errors.  Our final mean and median
velocity uncertainties were approximately 2 $\kms$.

By chance, 62 of our stars were also contained in the prior set of
Armandroff, Olszewski \& Pryor (1995, AOP).  These stars were in good
$\chi^2$ agreement with our own measurements, confirming the validity
of our uncertainty estimates.  Furthermore, the good agreement between
the two datasets argues against the earlier claim by AOP that Draco
possesses an extraordinarily large binary fraction because the excess
binaries would have inflated the $\chi^2$ disparity between the two
datasets more than was observed. In fact, the datasets are completely
consistent with a binary distribution and fraction comparable to those
in the solar neighbourhood.

These new data allow -- for the first time -- a robust measurement of
the variation of the velocity dispersion with projected radius.  Hints
of a variation have been noted by previous authors (Hargreaves et
al. 1996), but at much lower statistical significance. We have shown
that the velocity dispersion of Draco's stars increases with
increasing radius. This provides the first direct evidence for the
existence of an extended halo in any dSph.  The rise of the velocity
dispersion is robust against changes in the assumed binary fraction,
measurement errors and the subtracted solid body rotation.

The new dataset warrants a more sophisticated approach to dynamical
modelling than has been customary before. We have used the two
parameter family of spherical models introduced by Wilkinson et
al. (2001, Paper I), for which the projected density provides a good
match to the starcount data. The first parameter is the velocity
anisotropy of the stars, the second parameter is the dark matter
content. The models project to the same surface density on the sky,
but the dark matter content can vary between the extremes of
mass-follows-light and an extended dark halo with a harmonic
core. Similarly, the velocity can vary from the circular orbit model
through isotropy to the radial orbit model. The fit implements our
novel maximum likelihood algorithm, which makes proper allowance for
the effects of measurement errors and contamination by binary stars in
the radial velocity dataset.

Draco is best fit by a slightly tangentially anisotropic orbital
distribution and with a halo that falls off somewhat more slowly than
a flat rotation curve model.  We are able to rule out both a
mass-follows-light model and an extended halo with harmonic core at
the $3\sigma$ significance level (subject only to the limitations of
our parameterised models).  These conclusions are largely independent
of the assumed binary fraction and are insensitive to the subtraction
of solid body rotation. This provides the first definitive evidence
that the mass-to-light ratio increases outwards and that the dark
matter in Draco resides in a near-isothermal halo.

Our success in eliminating the systematic effects of the strong sky
lines in the region of the Ca triplet demonstrates that we have not
approached a fundamental limiting magnitude and that a 4m telescope
(like the WHT) should be capable of observing stars much fainter along
the giant branch in Draco.  The strong rise of the luminosity function
with magnitude implies that even a modest increase in limiting
magnitude will provide a much larger sample of stars for dynamical
modelling.  For instance, we estimate that extending the magnitude
limit to $V\sim 20.5$ will increase the sample size by a factor of
three or so. In other words, datasets of $\sim 600$ discrete radial
velocities in Draco are easily within the grasp of current technology.
We conclude that future prospects for mapping out the detailed
distribution of dark matter in Draco with stellar kinematics are
excellent.

\section*{Acknowledgments}
MIW and JK acknowledge financial support from PPARC, while NWE thanks
the Royal Society. We thank Chris Tout for helpful suggestions on the
modelling of giant branch binary evolution, and the anonymous referee
for useful comments and suggestions. We also thank Colin Frayn for his
help in acquiring the Draco velocity data.

\bye